\documentclass{aa}  
\usepackage{graphicx}
\usepackage{txfonts}
\begin{document}

   \title{Analysis of galaxy SEDs from far-UV to far-IR with CIGALE: 
          Studying a SINGS test sample} 

   \author{S. Noll\inst{1,2}
   \and    D. Burgarella\inst{2}
   \and    E. Giovannoli\inst{2}
   \and    V. Buat\inst{2}
   \and    D. Marcillac\inst{3}
   \and    J. C. Mu\~noz-Mateos\inst{4}}

   \offprints{S. Noll}

   \institute{Institut f\"ur Astro- und Teilchenphysik, Universit\"at
              Innsbruck, Technikerstr. 25/8, 6020 Innsbruck, Austria\\ 
              \email{Stefan.Noll@uibk.ac.at}
   \and       Observatoire Astronomique de Marseille-Provence, 38 rue 
              Fr\'ed\'eric Joliot-Curie, 13388 Marseille Cedex 13, France
   \and       Institut d'Astrophysique Spatiale, B\^at. 121, Universit\'e 
              Paris XI, 91405 Orsay, France              
   \and       Departamento de Astrof\'isica y Ciencias de la Atm\'osfera, 
              Universidad Complutense de Madrid, 28040 Madrid, Spain
}

   \date{Received / Accepted}

 
  \abstract
   {}
{Photometric data of galaxies covering the rest-frame wavelength range from
far-UV to far-IR make it possible to derive galaxy properties with a high
reliability by fitting the attenuated stellar emission and the related dust 
emission at the same time.}
{For this purpose we wrote the code CIGALE (Code Investigating GALaxy 
Emission) that uses model spectra composed of the Maraston (or PEGASE) 
stellar population models, synthetic attenuation functions based on a 
modified Calzetti law, spectral line templates, the Dale \& Helou dust 
emission models, and optional spectral templates of obscured AGN. Depending 
on the input redshifts, filter fluxes are computed for the model set and 
compared to the galaxy photometry by carrying out a Bayesian-like analysis.
CIGALE was tested by analysing 39 nearby galaxies selected from SINGS. The 
reliability of the different model parameters was evaluated by studying the 
resulting expectation values and their standard deviations in relation to the 
input model grid. Moreover, the influence of the filter set and the quality 
of photometric data on the code results was estimated.}
{For up to 17 filters with effective wavelengths between $0.15$ and 
$160$\,$\mu$m, we find robust results for the mass, star formation rate, 
effective age of the stellar population at 4000\,\AA{}, bolometric 
luminosity, luminosity absorbed by dust, and attenuation in the far-UV. 
Details of the star formation history (excepting the burst fraction) and the 
shape of the attenuation curve are difficult to investigate with the 
available broad-band UV and optical photometric data. A study of the mutual 
relations between the reliable properties confirms the dependence of star
formation activity on morphology in the local Universe and indicates a 
significant drop in this activity at about $10^{11}$\,M$_{\odot}$ towards 
higher total stellar masses. The dustiest galaxies in the SINGS sample are 
present in the same mass range. On the other hand, the far-UV attenuation of 
our sample galaxies does not appear to show a significant dependence on star
formation activity.}
{The results for our SINGS test sample demonstrate that CIGALE can be a 
valuable tool for studying basic properties of galaxies in the near and 
distant Universe if UV-to-IR data are available.}


   \maketitle
%

\section{Introduction}\label{introduction}

Apart from dark matter, galaxies consist of stars, gas, and dust. Stars 
produce the radiation that allows us to observe galaxies in a wide wavelength 
range. Interstellar gas mainly modifies the spectral energy distributions 
(SEDs) of galaxies by additional line emission or absorption. Finally, 
interstellar dust affects galaxy SEDs by extinction, i.e. absorption and 
scattering of stellar radiation, especially in the UV and optical and 
re-emission of the absorbed energy in the IR. The dust-induced energy 
conversion makes it difficult to study basic properties related to the 
stellar populations like the star formation rate (SFR), since a good 
knowledge of the galaxy SEDs over a wide wavelength range is necessary to 
evaluate the effect of dust.

In the past, the availability of IR galaxy data (IRAS and ISO) was mainly 
restricted to the nearby Universe. Hence, most low-to-high-redshift galaxies 
could be observed in the rest-frame UV to near-IR only. Since the total 
amount of dust emission in the IR could not be used to determine the amount 
of attenuation of the stellar continuum in the observed wavelength ranges, 
in particular, studies of broad-band SEDs suffered from difficulties to 
disentangle age, metallicity, and dust effects. The lack of information made 
it necessary to assume a typical shape of the attenuation law (Calzetti et 
al. \cite{CAL94}, \cite{CAL00}) and to apply simple recipes in the UV or 
optical to estimate the amount of attenuation. Possible variations in the 
galaxy dust properties could not be considered in this way, which caused high 
uncertainties in parameters such as the SFR. However, in view of many 
high-quality IR data collected by {\em Spitzer}, it is now possible to study 
SEDs of large galaxy samples up to high redshifts in a wide wavelength range, 
which allows us to better understand dust effects.       

In face of the complex SEDs of galaxies consisting of emission from different 
stellar populations which is modified by interstellar gas and dust in a
complicated way, it is clear that the derivation of galaxy properties needs
a realistic modelling of galaxy SEDs. For the stellar populations those 
models were produced by, e.g., Fioc \& Rocca-Volmerange (\cite{FIO97}), 
Bruzual \& Charlot (\cite{BRU03}), and Maraston (\cite{MARA05}). Dust 
attenuation curves were either derived by studying the SEDs of nearby 
star-forming galaxies (Calzetti et al. \cite{CAL94}, \cite{CAL00}) or by more 
theoretical radiative transfer models (e.g., Witt \& Gordon \cite{WIT00}). 
The emission of dust grains in the IR and polycyclic aromatic hydrocarbons 
(PAHs; see Peeters et al. \cite{PEE04} and references therein) in the mid-IR
was described by templates of Chary \& Elbaz (\cite{CHAR01}), Dale \& 
Helou (\cite{DAL02}), Lagache et al. (\cite{LAG03}, \cite{LAG04}), and 
Siebenmorgen \& Kr\"ugel (\cite{SIE07}). Models that also include stellar 
emission and dust attenuation at shorter wavelengths were published by, e.g., 
Silva et al. (\cite{SIL98}), Dopita et al. (\cite{DOP05}), and da Cunha 
et al. (\cite{CUN08}). The former also consider the evolution of dust 
properties dependent on the age of the stellar population. Sophisticated 
fitting codes of galaxy SEDs mainly focus on the derivation of redshifts and 
relatively robust galaxy properties like masses (e.g., Bolzonella et al. 
\cite{BOL00}; Feldmann et al. \cite{FEL06}; Walcher et al. \cite{WAL08}). 
Models for the IR are usually not included.

The above collection of publications on the modelling of galaxy SEDs shows 
that the comparison of data and models is an important tool for studying the
physical properties of galaxies. However, details of the listed models
like galaxy types, wavelength regimes, and characteristic parameters can 
differ a lot. On the other hand, the different astronomical questions and the 
available data sets require suitable models. Finally, optimised routines are
necessary in order to derive the characteristic galaxy properties under
investigation from the comparison of data and models. 

Hence, we feel that an observer-friendly fitting code for star-forming 
galaxies at different (given) redshifts is still missing which especially 
provides star formation histories, dust attenuation properties, and masses 
and uses photometric data covering the rest-frame UV to IR to constrain these 
properties. In other words, we are interested in a code that calculates the 
effect of dust on galaxy SEDs in a consistent way in order to obtain 
properties (such as the SFR or the effective age of the stellar populations) 
which are important for understanding the evolution of galaxies. Based on a 
procedure described in Burgarella et al. (\cite{BUR05}), we have developed a 
code that derives galaxy properties by means of a Bayesian-like analysis. In 
Sect.~\ref{code} we describe the models used and the fitting procedure. In 
Sect.~\ref{sings} we test the code for nearby galaxies selected from the 
SINGS sample (Kennicutt et al. \cite{KEN03}) for which good photometric data 
from the UV to the IR are available. Finally, the resulting findings 
concerning the applicability and effectivity of the code are discussed in 
Sect.~\ref{discussion}.

\section{The code}\label{code}

\begin{figure*}
\centering 
\includegraphics[height=8.8cm,clip=true]{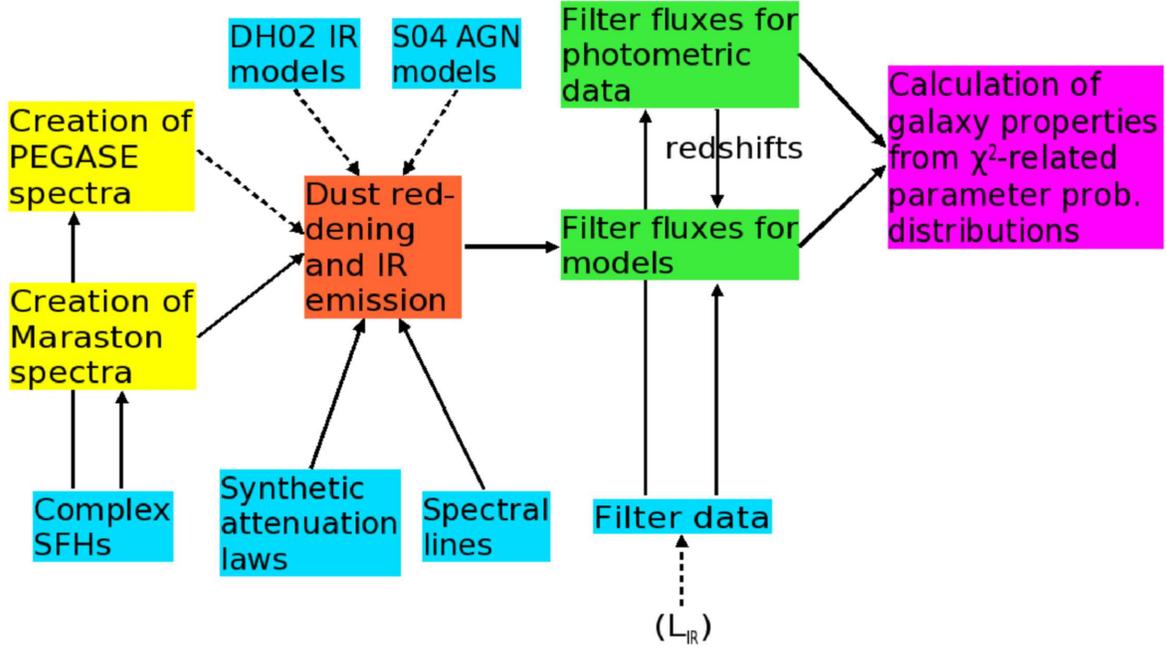}
\caption[]{Flow chart of {\em CIGALE}. The different programme modules and the
corresponding parameter and template inputs are shown from the left to the 
right. The alternative (and less preferred) use of the PEGASE models is 
marked by a dashed arrow. The other dashed arrows refer to the option to take 
either IR photometry directly (consideration of Dale \& Helou and possibly
Siebenmorgen et al. models) or externally estimated IR luminosities for 
considering the IR regime. The presented flow chart assumes that the object 
redshifts are taken from the photometric input catalogue.}
\label{fig_flowchart}
\end{figure*}

Our programme package {\em CIGALE}\footnote{Our code is provided at 
\texttt{http://www.oamp.fr/cigale/.}} ({\it C}ode {\it I}nvestigating 
{\it GAL}axy {\it E}mission) is characterised by a series of modules which 
are fed by spectra and parameter files. A flow chart of the code is shown in 
Fig.~\ref{fig_flowchart}. The general idea is to build stellar population 
models first. Secondly, the dust is considered by reddening the stellar SEDs 
and re-emitting the absorbed energy in the IR. Dust emission in the IR due to 
non-thermal sources can also be added. Interstellar lines are taken into 
account as corrections of the dust-affected SEDs. Redshift-dependent filter 
fluxes of the entire model set are, then, calculated for a direct comparison 
to the input object data. Finally, depending on the $\chi^2$ for the best-fit 
models of a set of bins in the parameter space, probability distributions as 
a function of the parameter value are calculated and used to derive weighted 
galaxy properties. Before we discuss this Bayesian-like analysis and the 
interpretation of its results in more detail in Sect.~\ref{fitting} and 
\ref{interpretation}, we start with a description of the models used in 
Sect.~\ref{models}.

\subsection{Models}\label{models}

For {\em CIGALE} we take models indicating the net emission from a galaxy in
the wavelength range from far-UV to far-IR. This means stellar emission, 
absorption and emission by dust, and at least the strongest interstellar 
absorption and emission lines have to be considered.

\subsubsection{Stars}\label{stars}

Concerning the stellar SEDs we have decided to focus on the models of 
Maraston (\cite{MARA05}), since they consider the thermally pulsating 
asymptotic giant branch (TP-AGB) stars\footnote{The Maraston models include 
TP-AGB spectra up to $2.5$\,$\mu$m only. Therefore, the flux level in a 
narrow wavelength range in the near-IR can still be systematically
underestimated.} in a realistic way. These stars are bright intermediate-age 
stars of $0.2$ to 1--2\,Gyr which mostly contribute from red optical to 
near-IR wavelengths. Therefore, they are particular important for a reliable 
stellar mass determination, but star-formation-related parameters are also 
affected. The insufficient consideration of TP-AGB stars in older but 
widespread models of Bruzual \& Charlot (\cite{BRU03}) and Fioc \& 
Rocca-Volmerange (\cite{FIO97}; PEGASE) typically increases the mass by 
$0.2$\,dex for star-forming galaxies with an important population of 
intermediate-age stars\footnote{For early-type galaxies in the nearby 
Universe the difference is expected to be lower, since most stars are 
distinctly older than the maximum age of a TP-AGB star.} (Maraston et al. 
\cite{MARA06}; Salimbeni et al. \cite{SALB09}). Therefore, {\em CIGALE} also 
allows PEGASE models to be used for backwards compatibility. 

For the calculation of complex stellar populations (CSPs) we consider single 
stellar populations (SSPs) of Maraston (\cite{MARA05}) and PEGASE of 
different metallicity $Z$ and Salpeter (\cite{SALP55}) or Kroupa 
(\cite{KRO01}) initial mass function (IMF). Masses based on the Salpeter IMF 
are about $0.2$\,dex higher than for the Kroupa IMF. The galaxy mass is 
normalised to 1\,M$_\odot$ and comprises the total mass of the stars (active 
and dead) plus gas that originates from stellar mass loss. This means that 
all SSP models start with a stellar mass fraction of 100\%. For a Salpeter 
IMF the fraction reaches about 70\% at the age of the Universe. The total 
stellar mass is provided by the code. For obtaining CSPs the SSPs of 
different ages are weighted and added according to the star formation 
scenario chosen. We provide ``box models'' with constant star formation over 
a limited period and ``$\tau$ models'' with exponentially decreasing SFR, 
respectively. While for the former scenario and ongoing star formation the 
instantaneous SFR at look-back time $t' = 0$ is simply calculated by the 
galaxy mass divided by the age, i.e. $M_\mathrm{gal}/t$, the SFR of the 
latter scenario results in  
\begin{equation}
\mathrm{SFR} = \frac{M_\mathrm{gal}}{\tau (e^{t / \tau} - 1)},
\end{equation} 
where $\tau$ is the $e$-folding or decay time. We focus on these $\tau$ 
models for the rest of the paper. 
  
\begin{figure}
\centering 
\includegraphics[width=8.8cm,clip=true]{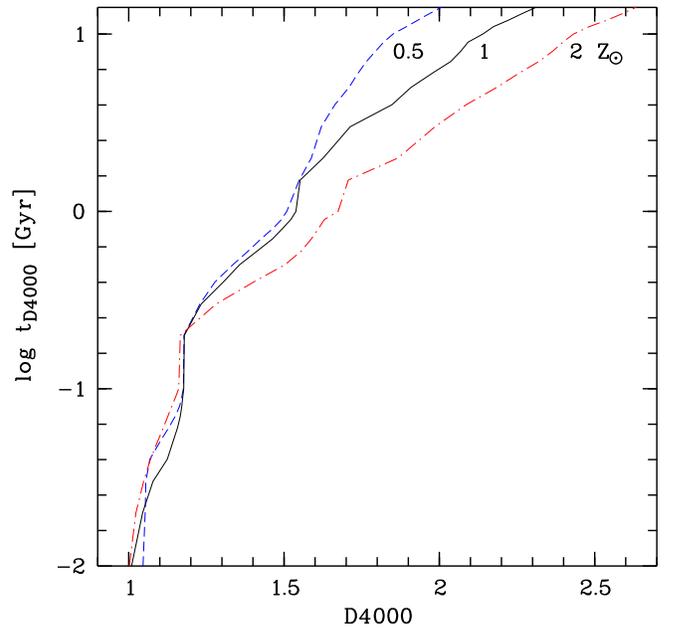}
\caption[]{Relation between the age $t_{\,\mathrm{D4000}}$ and the strength 
of the D4000 break for Maraston (\cite{MARA05}) single stellar population 
models with Salpeter IMF and metallicities 0.5, 1, and 2\,Z$_\odot$. Local, 
low amplitude extrema in the curves were smoothed out to avoid any 
ambiguities.}
\label{fig_age_d4000}
\end{figure}

{\em CIGALE} allows two CSP models to be combined. Both model components are, 
then, linked by their mass fraction. This scenario makes it possible to 
consider bursts on top of an older passive or more quiescent star-forming 
stellar population. Although this approach allows galaxy SEDs to be fit quite 
well, it is clear that this scenario can only be a rough representation of 
the real star formation history (SFH) of a galaxy. On the other hand, the 
more complex SFH scenarios are, the more SED fitting suffers from 
degeneracies. In view of the uncertainties in the basic stellar population 
model parameters, {\em CIGALE} also computes two effective parameters: the 
mass-weighted age 
\begin{equation}
t_{\,\mathrm{M}} = \frac{\int_{\,0}^{\,t} t' \mathrm{SFR}(t') \,\mathrm{d}t'}
{\int_{\,0}^{\,t} \mathrm{SFR}(t') \,\mathrm{d}t'}
\end{equation}
and the age $t_{\,\mathrm{D4000}}$ derived from the D4000 
break\footnote{ratio of the average flux per frequency unit of the wavelength 
ranges 4000--4100\,\AA{} and 3850--3950\,\AA{}} (see Balogh et al. 
\cite{BALO99}) of the unreddened SED for a single stellar population, which 
shows a quasi-monotonic increase of D4000 with age (see 
Fig.~\ref{fig_age_d4000}). The two ages are complementary, since they trace 
the old mass-dominating and the young light-dominating stellar population, 
respectively.      

\subsubsection{Dust attenuation}\label{attenuation}

\begin{figure}
\centering 
\includegraphics[width=8.8cm,clip=true]{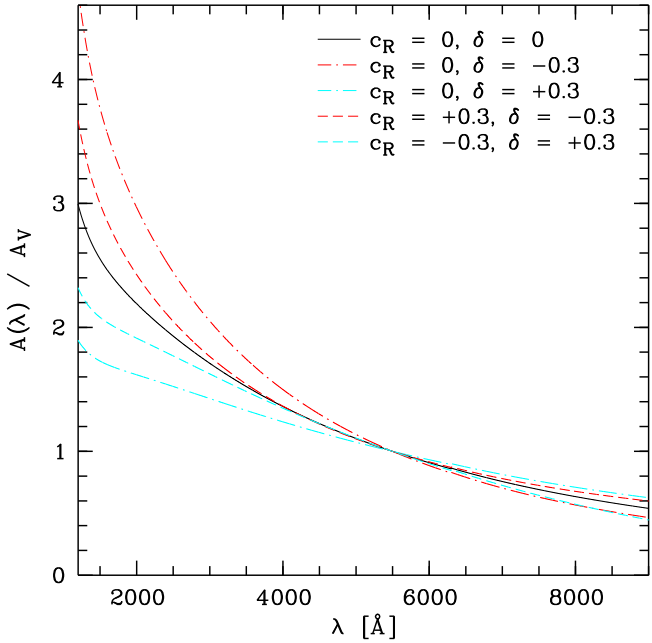}
\caption[]{Illustration of synthetic attenuation laws differing in 
$c_\mathrm{R}$ and $\delta$. Apart from a pure Calzetti law 
($c_\mathrm{R} = 0$ and $\delta = 0$; solid curve), Calzetti laws modified by 
multiplication of a power law are shown. Lower $\delta$ cause steeper slopes 
in the UV. In two cases the change of $R_V$ by the power law is compensated 
by $c_\mathrm{R} = - \delta$ (dashed curves).}
\label{fig_illattcurv}
\end{figure}

The absorption and scattering of star light by dust is described by 
extinction curves. In the case of galaxies these laws are preferentially 
called attenuation curves because of the dust-induced re-scattering of 
radiation in direction towards the observer, which is negligible if only 
a single star without a circumstellar dust shell is taken into account (see 
Kr\"ugel \cite{KRU09}). For typical sightlines towards Milky Way stars the 
extinction curve is characterised by increasing extinction at shorter 
wavelengths, which causes a reddening of the spectra, and the so-called UV 
bump at about 2175\,\AA{}, which produces a broad absorption feature (e.g., 
Stecher \cite{STE69}; Cardelli et al. \cite{CAR89}; Fitzpatrick \& Massa 
\cite{FIT07}). For typical sightlines towards the Large and Small Magellanic 
Cloud (LMC and SMC) the extinction curves are steeper and the UV bumps are 
weaker than in the Milky Way (Gordon et al. \cite{GOR03}). The most extreme 
curve is found for the SMC, where the 2175\,\AA{} feature is almost vanished 
completely. A non-existing UV bump and a moderate far-UV rise characterises 
the effective, average attenuation curve of nearby starburst galaxies found 
by Calzetti et al. (\cite{CAL94}, \cite{CAL00}). However, about 30\% of the 
UV-luminous star-forming galaxies at $1 < z < 2.5$ seem to exhibit a 
significant 2175\,\AA{} feature which can be as strong as those found in the 
LMC (Noll \& Pierini \cite{NOL05}; Noll et al. \cite{NOL07}, \cite{NOL09}). 
Moreover, the typical width of the UV bump of these galaxies is only about 
60\% of those measured for characteristic Milky Way and LMC sightlines. 

These observations show that a single attenuation curve as given by Calzetti 
et al. (\cite{CAL00}) and often used in the literature is probably 
insufficient to describe star-forming galaxies, in particular, at higher 
redshifts. Hence, we apply a more complex ansatz to describe the attenuation 
in our model galaxies. Since the Calzetti law seems to provide reasonable
amounts of attenuation in starburst galaxies at least as an average for large
samples, we take this frequently used curve as basis. For wavelengths below 
1200\,\AA{}, for which the Calzetti law is not defined, we linearly 
extrapolate by using the value and slope at 1200\,\AA{}. This approach, which
resembles the method of Bolzonella et al. (\cite{BOL00}), avoids
extreme slopes at low wavelengths that are not supported by observations
(Leitherer et al. \cite{LEI02}). 

At first, the Calzetti law $k(\lambda) + R_{V,\,0}$ 
($R_{V,\,0} = 4.05 \pm 0.8$) is modified by adding a UV bump which is modelled 
by a Lorentzian-like ``Drude'' profile 
\begin{equation}
D_{\lambda_0,\,\gamma,\,E_\mathrm{bump}}(\lambda) = 
\frac{E_\mathrm{bump}\,\lambda^2\gamma^2}{(\lambda^2 - {\lambda_0}^2)^2 + 
\lambda^2\gamma^2},
\end{equation}  
where $\lambda_0$, $\gamma$, and $E_\mathrm{bump}$ are the central wavelength, 
width ($\approx$~FWHM), and amplitude (or maximum height), respectively
(Fitzpatrick \& Massa \cite{FIT90}, \cite{FIT07}; Noll et al. \cite{NOL09}). 
In order to produce different slopes, the modified Calzetti law is multiplied 
by a power law $(\lambda/\lambda_V)^{\,\delta}$, where 
$\lambda_V = 5500$\,\AA{} is the reference wavelength of the $V$ filter. This 
choice allows the slope to be changed without altering the scaling parameter 
visual attenuation $A_V$. Slopes $\delta > 0$ represent flatter slopes than 
the Calzetti law, while $\delta = 0$ exactly reproduces this basic curve if 
no bump is considered (see Fig.~\ref{fig_illattcurv}). If the resulting curve 
gives a negative attenuation correction, which even happens for a pure 
Calzetti law in the near-IR, no correction is applied. Since $\delta \ne 0$ 
causes a change in the original scaling parameter of the Calzetti attenuation 
curve $E_{B-V}$, the real $R_V = A_V / E_{B-V}$ differs from the value 
$R_{V,\,0} = 4.05$ of the Calzetti law. In order to compensate for this 
effect and to have more freedom in the choice of this constant, the 
attenuation law is multiplied by the factor 
$(1 - a_\mathrm{R} \, c_\mathrm{R}) / R_{V,\,0}$. For $a_\mathrm{R} = 1.12$ 
this term guarantees roughly constant $R_V$ if 
$c_\mathrm{R} + \delta = \mathrm{const}$ (see Fig.~\ref{fig_illattcurv}). For 
reasonable slope corrections $-0.3 < \delta < 0.3$ the real $R_V$ changes by 
about +1\% or less. In summary, the dust attenuation in magnitudes as 
function of wavelength is
\begin{equation}
A(\lambda) = A_V 
\Big[ (k(\lambda) + D_{\lambda_0,\,\gamma,\,E_\mathrm{bump}}(\lambda)) 
\frac{1 - 1.12 \, c_\mathrm{R}}{4.05} + 1 \Big] 
\Big(\frac{\lambda}{5500\,\mathring{\mathrm{A}}}\Big)^{\delta}.
\end{equation}

The attenuation correction is applied to both possible stellar population
components (see Sect.~\ref{stars}) individually by allowing for different
$A_V$. In practice, the visual attenuation of the young $\tau$ model 
$A_{V,\,\mathrm{ySP}}$ and a reduction factor of the attenuation for the old
$\tau$ model $f_\mathrm{att}$ are used. This simple procedure takes into 
account that stars of different age can be affected by different amounts of 
dust (Silva et al. \cite{SIL98}; Pierini et al. \cite{PIE04}; Panuzzo et al. 
\cite{PAN07}; Noll et al. \cite{NOL07}).  

In view of the complexity of the attenuation curves and the dependence of the
total attenuation on the SFH due to the introduction of $f_\mathrm{att}$, we 
characterise the effective obscuration of the stellar radiation by two 
additional parameters that are derived from the final model SEDs. 
$A_\mathrm{FUV}$ and $A_V$ are defined as the effective attenuation factors 
in magnitudes at $1500 \pm 100$~\AA{} and $5500 \pm 100$~\AA{}, respectively.
Both parameters are complementary. While $A_\mathrm{FUV}$ probes the dust
obscuration of the young stellar population only, $A_V$ also traces the 
attenuation of older and cooler stars. The latter parameter equals 
$A_{V,\,\mathrm{ySP}}$ if only one $\tau$ model is used to describe the SFH 
of a galaxy.

\subsubsection{Dust emission}\label{dustemission}

\begin{figure}
\centering 
\includegraphics[width=8.8cm,clip=true]{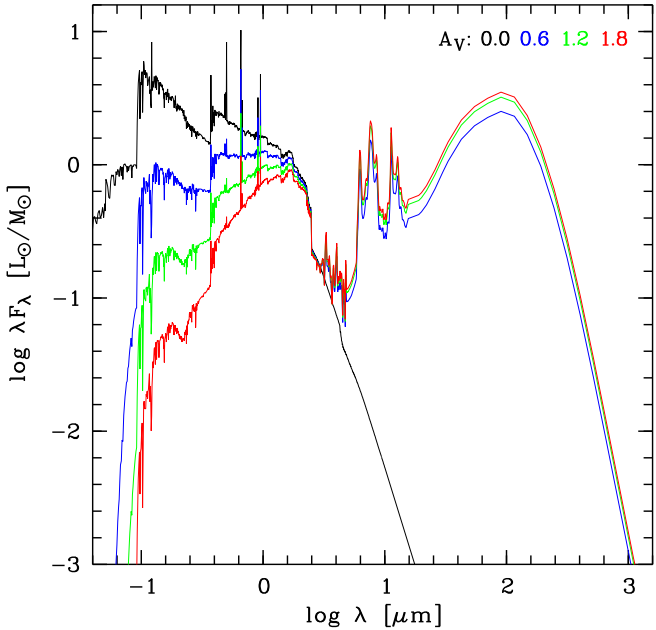}
\caption[]{Illustration of a set of complete models differing in $A_V$ (see 
legend). The higher $A_V$ is, the lower the flux in the UV and the higher the 
flux in the IR is. All models are characterised by solar metallicity, 
$\tau = 10$\,Gyr, $t = 1$\,Gyr, Calzetti law (i.e. $c_\mathrm{R} = 0$ and 
$\delta = 0$), an LMC-like UV bump, no AGN contamination, and $\alpha = 2.0$.}
\label{fig_mod_ebv}
\end{figure}

\begin{figure}
\centering 
\includegraphics[width=8.8cm,clip=true]{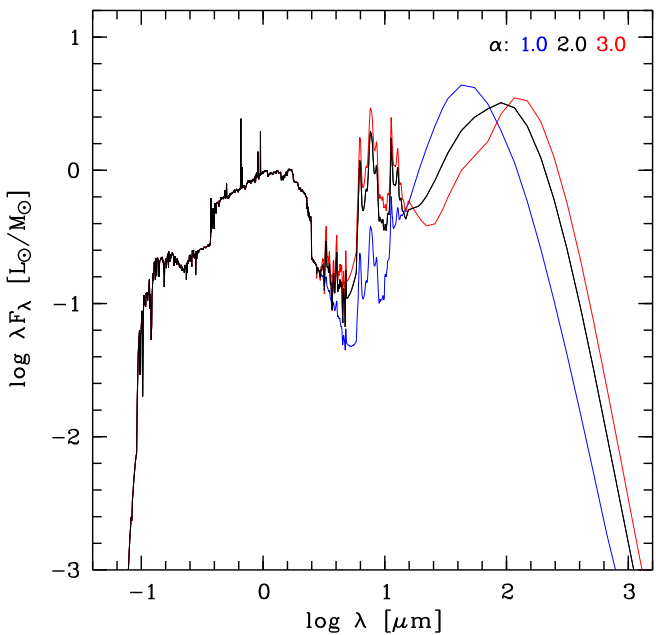}
\caption[]{Illustration of a set of complete models differing in $\alpha$
(see legend). The higher $\alpha$ is, the larger the peak wavelength of the
dust emission is. All models are characterised by solar metallicity, 
$\tau = 10$\,Gyr, $t = 1$\,Gyr, Calzetti law (i.e. $c_\mathrm{R} = 0$ and 
$\delta = 0$), an LMC-like UV bump, no AGN contamination, and 
$A_V = 1.2$\,mag.}
\label{fig_mod_alpha}
\end{figure}

Dust re-radiates energy predominantly absorbed in the UV and optical at 
relatively long, mid-IR to sub-mm wavelengths. In the far-IR dust emission 
is characterised by continuum radiation that originates from dust grains in 
thermal equilibrium. The wavelength distribution of the emission depends on 
grain temperature, i.e. longer wavelengths indicate lower temperatures. In 
the mid-IR the continuum radiation of star-forming galaxies is mainly caused 
by stochastically-heated\footnote{Thermal emission of very small grains in the
mid-IR is restricted to extremely intense heating environments that are not 
typical of normal star-forming galaxies (e.g., Dale et al. \cite{DAL01}).} 
very small grains (see, e.g., da Cunha et al. \cite{CUN08} and references 
therein). This emission is superimposed by prominent broad emission features 
of polycyclic aromatic hydrocarbon (PAH) molecules\footnote{Very small 
grains and especially PAHs in their low energy states may significantly 
contribute to the emission at mm and cm wavelengths. In particular, spinning
PAHs are proposed as the origin of the ``anomalous emission'' at these 
wavelengths (see Ysard \& Verstraete \cite{YSA09}).} (see Puget \& L\'eger 
\cite{PUG89}; Sturm et al. \cite{STU00}; Draine \cite{DRA03}; Peeters et al. 
\cite{PEE04}). In contrast, mid-IR spectra dominated by hot dust continuum 
emission and lacking significant PAH emission are usually related to active 
galactic nuclei (AGN).
 
The IR properties of star-forming galaxies can be described by physically
motivated multi-parameter models (e.g., Dopita et al. \cite{DOP05}; 
Siebenmorgen \& Kr\"ugel \cite{SIE07}; da Cunha et al. \cite{CUN08}). 
However, since the IR data of galaxies at higher redshifts are usually quite
sparse, we have decided to rely on semi-empirical, one-parameter models 
(Chary \& Elbaz \cite{CHAR01}, Dale \& Helou \cite{DAL02}, Lagache et al. 
\cite{LAG03}, \cite{LAG04}). For {\em CIGALE} we take the 64 templates of 
Dale \& Helou (\cite{DAL02}), which are parametrised by the power law slope
of the dust mass distribution over heating intensity $\alpha$. For higher
$\alpha$ the contribution of relatively quiescent galactic regions 
characterised by weak radiation fields becomes more important and the dust 
emission peaks at longer wavelengths (see Dale et al. \cite{DAL01}). The PAH 
emission pattern of the Dale \& Helou (\cite{DAL02}) models shows only little 
variation. This agrees with the observations as long as the IR emission is
not significantly affected by an AGN (e.g., Peeters et al. \cite{PEE04}). As 
for the Chary \& Elbaz (\cite{CHAR01}) and Lagache et al. (\cite{LAG03}) 
models, a characteristic IR luminosity $L_\mathrm{IR}$ can be assigned to 
each Dale \& Helou (\cite{DAL02}) template. Those calibrations for 
luminosity-dependent IR SEDs of the Dale \& Helou models are available from 
Chapman et al. (\cite{CHAP03}) and Marcillac et al. (\cite{MARC06}). Since 
individual galaxies can significantly differ from these relations for sample 
averages, we do not directly use them for {\em CIGALE}. However, they can be 
taken to adapt the $\alpha$ parameter space for the kind of objects 
investigated.    

The semi-empirical Dale \& Helou templates include stellar emission in the 
near-IR range and below (see Dale et al. \cite{DAL01}). In order to combine 
them with stellar population models the stellar contribution needs to be 
subtracted. Therefore, we scaled a 5\,Gyr-old passively-evolving Maraston 
model to the flux at $2.5 - 3$\,$\mu$m in the different Dale \& Helou models 
to remove the stellar emission. For wavelengths below $2.5$\,$\mu$m we set 
the flux in the IR models to zero, in any case. The flux reduction drops 
below 50\% at about 5\,$\mu$m. The choice of the stellar population model is 
not critical, since the slope of the stellar continuum at these long 
wavelengths shows only little variation for different SFHs. 
 
The IR templates are linked to the attenuated stellar population models by 
the dust luminosity $L_\mathrm{dust}$, i.e. the luminosity ``absorbed'' by 
the dust and re-emitted in the IR. Hence, the scaling of the Dale \& Helou 
templates is not a free parameter. The change of the luminosity transfer with 
increasing $A_V$ for a fixed attenuation law is illustrated in 
Fig.~\ref{fig_mod_ebv}. On the other hand, Fig.~\ref{fig_mod_alpha} 
demonstrates for fixed $L_\mathrm{dust}$ the change in the shape of the IR 
SED if the only IR-specific parameter $\alpha$ is modified.

The equality between the dust-absorbed and dust-emitted luminosity can be 
violated by dust emission caused by a non-thermal source. In particular, 
highly dust-enshrouded AGN represent a problem, since they are difficult to
identify in the UV and optical. Consequently, galaxies with such an AGN 
contribution look like normal star-forming galaxies in the far-UV-to-far-IR
wavelength range excepting a warm/hot dust emission in the IR. Therefore,
we allow for an additional IR dust emission component which is not balanced 
by dust absorption of stellar emission at shorter wavelengths.

Siebenmorgen et al. (\cite{SIE04a}, \cite{SIE04b}) provide almost 1500 AGN
models differing in the luminosity of the non-thermal source, the outer 
radius of a spherical dust cloud covering the AGN, and the amount of 
attenuation in the visual caused by the cloud. In principle, all models can
be fed into {\em CIGALE}. Focusing on SEDs providing PAH-free mid-IR emission, 
the number of suitable models significantly decreases, however. As reference 
model we take $L = 10^{12}$\,L$_\odot$, $R = 125$\,pc, and $A_V = 32$\,mag. 
Using this or similar models allows us to disentangle IR dust emission 
components caused by stellar and AGN radiation, respectively, since the 
AGN-related component peaks at significantly shorter wavelengths than the 
stellar one. Admittedly, the real SEDs of obscured AGN could significantly 
deviate from our preferred model. However, this is almost impossible to test 
with broad-band photometric data only.

\subsubsection{Gas}\label{gas}

Strong spectral lines or blends can easily change filter-averaged fluxes by 
10\% or more. Therefore, spectral features need to be considered to avoid
systematic deviations in the spectra. The crucial features are nebular 
emission lines of, e.g., H\,I, [O\,II], and [O\,III] in the optical and 
interstellar absorption lines of different ions in the UV. The UV is also 
affected by stellar wind lines like C\,IV and nebular emission, i.e.
essentially the strongly-varying Ly$\alpha$ line. The narrow nebular emission 
lines in the IR are not explicitly considered, since we assume that they 
already contribute to the low-resolution semi-empirical templates of Dale \& 
Helou (\cite{DAL02}).   

The spectral line correction is performed by using empirical line templates. 
We have taken the Kinney et al. (\cite{KIN96}) starburst spectra to derive 
two optical emission line templates mainly differing in the strength of the 
oxygen lines. For the UV we have derived two line templates from composites 
of high-redshift galaxies of Noll et al. (\cite{NOL04}), since their UV
spectra have a better quality than those of Kinney et al. (\cite{KIN96}).
The UV line templates include all lines between Lyman limit and 3000\,\AA{}
excepting those which are already present in the stellar population models.
The selection of the optical and UV templates depends on the flux ratio of 
the wavelength ranges $1420 - 1490$\,\AA{} and $3400 - 3600$\,\AA{} in the
attenuated stellar population models in order to reproduce the change of
line strengths with UV continuum reddening in star-forming galaxies. The
best agreement with the observed spectra is achieved if for ratios $\ge 3.7$ 
the templates with strong emission and weak absorption features are 
taken. For ratios $< 1.0$ and star-related 
$L_\mathrm{dust}/L_\mathrm{bol} < 0.1$ the spectral line templates are not 
considered at all in order to reproduce mainly passively-evolving galaxies.
Before the selected line templates are added to the models, the optical line 
templates are scaled to the average flux in the wavelength range 
$3400 - 3600$\,\AA{}. On the other hand, the UV line templates are, first, 
multiplied to the dust-attenuated models cleaned from stellar lines by 
interpolation. The effect of the two sets of UV and optical line templates on 
model spectra is visible in Fig.~\ref{fig_mod_ebv}. 

The described procedure for the consideration of interstellar lines in the 
model SEDs is, of course, relatively simple. However, it avoids systematic 
errors, although with significant uncertainties for individual objects with 
SEDs deviating from the composites analysed. A clear advantage of this 
phenomenological approach is certainly the avoiding of new free parameters in 
the code.

\subsection{Fitting and parameter probability distributions}\label{fitting}

\begin{figure}
\centering 
\includegraphics[width=8.8cm,clip=true]{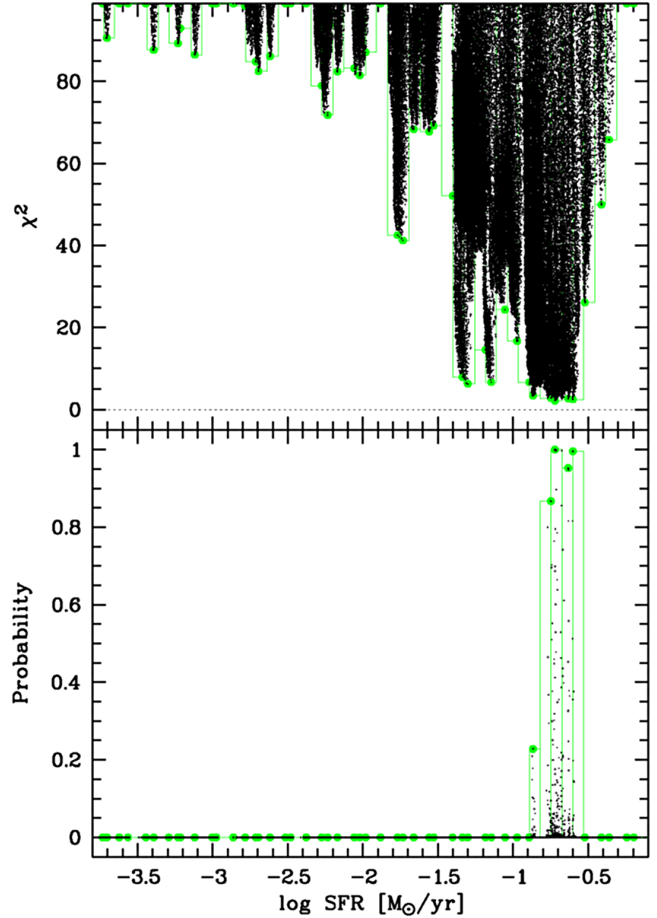}
\caption[]{PDF derivation for the SFR of a test galaxy. About 
$7 \times 10^5$~models are shown (small dots). The parameter space is divided 
into 50 bins, which are indicated by histograms. The best model of each bin 
is marked by a filled circle. The upper panel exhibits the $\chi^2$ 
distribution as a function of SFR. The lower panel shows the corresponding 
probabilities for the normalised $\chi^2$, i.e. 
$\chi^2 - \chi^2_\mathrm{best}$. Expectation values and standard deviations 
for the SFR can be derived from the probability distribution of the 
bin-specific best-fit models.}
\label{fig_illppd_SFR}
\end{figure}

\begin{figure}
\centering 
\includegraphics[width=8.8cm,clip=true]{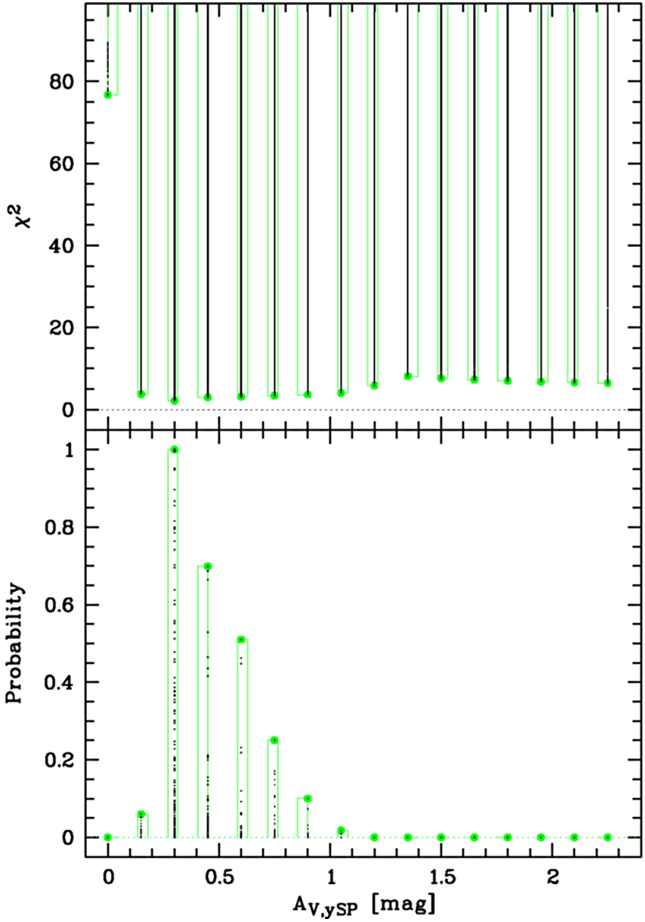}
\caption[]{PDF derivation for $A_{V,\,\mathrm{ySP}}$. In this example, the 
discrete model $A_{V,\,\mathrm{ySP}}$ values increase in steps of $0.15$\,mag. 
For the description of the plot see Fig.~\ref{fig_illppd_SFR}.}
\label{fig_illppd_A_V}
\end{figure}

The observational input data of {\em CIGALE} are photometric filter fluxes. 
Hence, the filter fluxes for the entire set of models are derived for
the known redshifts of the objects given. The final models taken also include 
a correction of the redshift-dependent absorption of the intergalactic medium
(IGM) shortwards of Ly$\alpha$. Here we take the more recent algorithm of
Meiksin (\cite{MEI06}) instead of the more frequently used corrections of 
Madau (\cite{MAD95}).       

For sparse IR data it is possible to provide an externally computed IR 
luminosity to the code instead of using the IR filters directly. For example,
the Dale \& Helou (\cite{DAL02}) models can be used in combination with the 
calibrations of Chapman et al. (\cite{CHAP03}) and Marcillac et al. 
(\cite{MARC06}) to derive typical $L_\mathrm{IR}$ from MIPS 24\,$\mu$m flux
densities. In this case, $L_\mathrm{IR}$ is converted into flux units for an 
artificial filter centred at 100\,$\mu$m. This is, then, compared to the dust 
luminosity $L_\mathrm{dust}$ derived for each model converted to a filter 
flux as well. In order to save computing time no IR models need to be 
considered in this case.  

The comparison of model and noise-affected object photometry\footnote{Since
filter-averaged fluxes are compared, the calibration of possibly considered 
{\em Spitzer} MIPS flux densities has to be changed to 
$\nu f_\nu = \mathrm{const}$. Consequently, the observed flux densities at 
24, 70, and 160\,$\mu$m have to be multiplied by $1.040$, $1.070$, and 
$1.043$, respectively (cf. MIPS data handbook), before they can be used in 
the code.}, i.e. of $f_{\mathrm{mod},i}$ (per $M_\odot$) and 
$f_{\mathrm{obs},i}$ for $k$ filters, is carried out for each model by the 
minimisation of 
\begin{equation}\label{eq_chi2}
\chi^2(M_\mathrm{gal}) = \sum_{i=1}^{k} 
\frac{(M_\mathrm{gal} f_{\mathrm{mod},i} - f_{\mathrm{obs},i})^2}
{\sigma_{\mathrm{obs},i}^2}
\end{equation}
with the galaxy mass $M_\mathrm{gal}$ (in $M_\odot$) as a free parameter. The 
statistical photometric errors are considered by $\sigma_{\mathrm{obs},i}$. 
The resulting minimum $\chi^2$ of each model can be compared to determine the 
model showing the best $\chi^2$ of the entire model grid for the object SED 
investigated. Moreover, the model-related $\chi^2$ allow us to perform a more 
sophisticated analysis to obtain galaxy properties based on probability 
distribution functions (PDFs). 

Probabilities for individual models are often computed using the exponential
term $e^{-\chi^2/2}$ (e.g., Kauffmann et al. \cite{KAU03a}; Salim et al. 
\cite{SALI07}; Walcher et al. \cite{WAL08}). Then, PDFs for each parameter
can be derived by calculating the probability sums of the models in given 
bins of the parameter space. However, the results of this `sum' method depend 
on the model density in the parameter space. In the case of a bad choice of 
the input model parameter values, this can cause an unintentional bias. 
Although the introduction of a lower threshold probability for the 
consideration of models can alleviate this effect, we adopt a different 
approach based on the best-fit models for given bins in the parameter 
space and integrated probabilities, which we describe in the following. We 
refer to this approach as `max' method. We compare the results of both 
methods in Sect.~\ref{fitquality}.     

The likelihood of a model can be inferred from $\chi^2$ by integrating the
corresponding probability density function from $\chi^2$ to infinity:
\begin{equation}
p(\chi^2,n) = \int_{\chi^2}^{\,\infty} \frac{x^{(n/2)-1}e^{-x/2}}
{2^{n/2}\, \Gamma(n/2)}\, \mathrm{d}x.
\end{equation}  
$\Gamma$ denotes the Gamma function and $n = k - 1$ the filter-related 
degrees of freedom. Since the minimum $\chi^2$, i.e. the fit quality, can 
vary a lot for an object sample, we prefer to use normalised 
$\chi^2_\mathrm{norm} = \chi^2 - \chi^2_\mathrm{best}$ (although not forced 
by the code) for computing $p$ in order to guarantee similar (and better 
comparable) probability distributions in the model sets for the different 
objects. This means that the best-fit model always has $p = 1$.   

For calculating the object-related expectation values and uncertainties for
the different model parameters, PDFs $P(x)$ depending on the parameter value 
$x$ are necessary. This requires to link the $p(\chi^2,n)$ for all $m$ models 
studied to the $P(x)$. We do so by introducing a fixed number $b$ of 
equally-sized bins for each parameter. The range of bins is delimited by the 
lowest and highest value of a parameter in the model set. We now derive the 
characteristic probability $P_i$ of each bin $i$ by searching the maximum 
$p_j(\chi^2,n)$ of the models $j$ located inside the bin $i$. The prior 
$a_{ji} = 1$ if a model $j$ belongs to the bin $i$, otherwise $a_{ji} = 0$. 
In summary, this procedure can be written in a Bayesian-like style:  
\begin{equation}   
P_i(x) = \max_{j=1,...,m}(p_j(\chi^2,n)\, a_{ji}).
\end{equation}
The resulting PDF $P_i(x)$ envelops the distribution of models in the 
$x$--$p$ plane (see Fig.~\ref{fig_illppd_SFR} for a continuum of $x$ and 
Fig.~\ref{fig_illppd_A_V} for discrete $x$). The already mentioned advantage 
of this approach is that $P_i(x)$ does not depend on the model density as a 
function of $x$. Consequently, the arbitrary choice of values for a parameter 
can hardly change $P_i(x)$ as long as the covered range of $x$ is comparable.

Taking the $P_i(x)$ as weights for each bin, the expectation value of each
parameter is given as
\begin{equation}   
\langle x \rangle = \frac{\sum_{i=1}^{b} P_i x_i}{\sum_{i=1}^{b} P_i}.
\end{equation}
For $x_i$ we directly use the parameter value of the best-fit model of each 
bin. It corresponds to the mean value of all models in a given bin if the 
number of bins is significantly higher than the number of realised parameter 
values, which should be the case for most model parameters. Finally, the 
standard deviation is derived by  
\begin{equation}   
\sigma_x = \sqrt{\frac{\sum_{i=1}^{b} P_i (x_i - \langle x \rangle)^2}
{\sum_{i=1}^{b} P_i}}.
\end{equation}

\begin{table}
\caption[]{Description of the output parameters of {\em CIGALE}}     
\label{tab_pardesc}
\centering
\begin{tabular}{l l l l}
\hline\hline
\noalign{\smallskip}
Par. & Unit & Description & Dependencies \\
\noalign{\smallskip}
\hline
\noalign{\smallskip}
$Z$ & --- & metallicity ($\mathrm{Z}_\odot = 0.02$) & basic \\
$\tau_\mathrm{oSP}$ & Gyr & $e$-folding time for & basic \\
& & old SP model & \\
$t_{\,\mathrm{oSP}}$ & Gyr & age of old SP model & basic \\
$\tau_\mathrm{ySP}$ & Gyr & $e$-folding time for & basic \\
& & young SP model & \\
$t_{\,\mathrm{ySP}}$ & Gyr & age of young SP model & basic \\
$f_\mathrm{burst}$ & --- & mass fraction of & basic \\
& & young SP model & \\
\noalign{\smallskip}
$t_{\,\mathrm{M}}$ & Gyr & mass-weighted age & dep. on SP par. \\ 
$t_{\,\mathrm{D4000}}$ & Gyr & D4000-related age & dep. on SP par. \\
\noalign{\smallskip}
$\lambda_0$ & \AA{} & central wavelength of & basic \\
& & UV bump & \\
$\gamma$ & \AA{} & FWHM of UV bump & basic \\
$E_\mathrm{bump}$ & --- & amplitude of UV bump & basic \\ 
$c_\mathrm{R}$ & --- & change of $R_V$ & basic \\ 
$\delta$ & --- & slope correction of & basic \\
& & Calzetti law & \\
$A_{V,\,\mathrm{ySP}}$ & mag & $V$-band attenuation & basic \\
& & for young SP model & \\ 
$f_\mathrm{att}$ & --- & reduction of $A_V$ & basic \\
& & for old SP model & \\
\noalign{\smallskip}
$A_\mathrm{FUV}$ & mag & attenuation at 1500\,\AA{} & dep. on SP and \\
& & & attenuation par. \\
$A_V$ & mag & attenuation at 5500\,\AA{} & dep. on SP and \\
& & & attenuation par. \\
\noalign{\smallskip}
$\alpha$ & --- & slope of IR model & basic \\
$N_\mathrm{AGN}$ & --- & number of AGN model & basic \\
$f_\mathrm{AGN}$ & --- & $L_\mathrm{dust}$ fraction of & basic \\ 
& & AGN model & \\
\noalign{\smallskip}
$N_\mathrm{emlin}$ & --- & template number for & dep. on SP and \\
& & optical lines & attenuation par. \\
$N_\mathrm{abslin}$ & --- & template number for & dep. on SP and \\
& & UV lines & attenuation par. \\
\noalign{\smallskip}
$M_\mathrm{gal}$ & M$_\odot$ & galaxy mass & scaling par. \\
\noalign{\smallskip}
$M_\mathrm{star}$ & M$_\odot$ & total stellar mass & dep. on SP par. \\
& & & and $M_\mathrm{gal}$ \\
SFR & M$_\odot$/yr & instantaneous SFR & dep. on SP par. \\
& & & and $M_\mathrm{gal}$ \\
$L_\mathrm{bol}$ & L$_\odot$ & bolometric luminosity & dep. on all basic \\
& & & par. and $M_\mathrm{gal}$ \\
$L_\mathrm{dust}$ & L$_\odot$ & dust luminosity & dep. on all basic \\
& & & par. and $M_\mathrm{gal}$ \\
\noalign{\smallskip}
\hline
\end{tabular}
\end{table}

{\em CIGALE} provides results for 16 basic input parameters\footnote{The 
true number of simultaneously analysed, basic parameters is lower than 16 and 
should not exceed 10 in most cases, since parameters such as the metallicity 
or the central wavelength of the UV bump are usually defined by a single 
value.}, the scaling parameter $M_\mathrm{gal}$, and 10 additional output 
parameters that depend on different basic model properties. A complete list 
is given in Table~\ref{tab_pardesc}. Apart from mean value and error, the 
code outputs the parameter values of the best-fit model and the full PDFs for 
each parameter if desired.

\subsection{Interpretation of the code results}\label{interpretation}

\begin{figure}
\centering 
\includegraphics[width=8.8cm,clip=true]{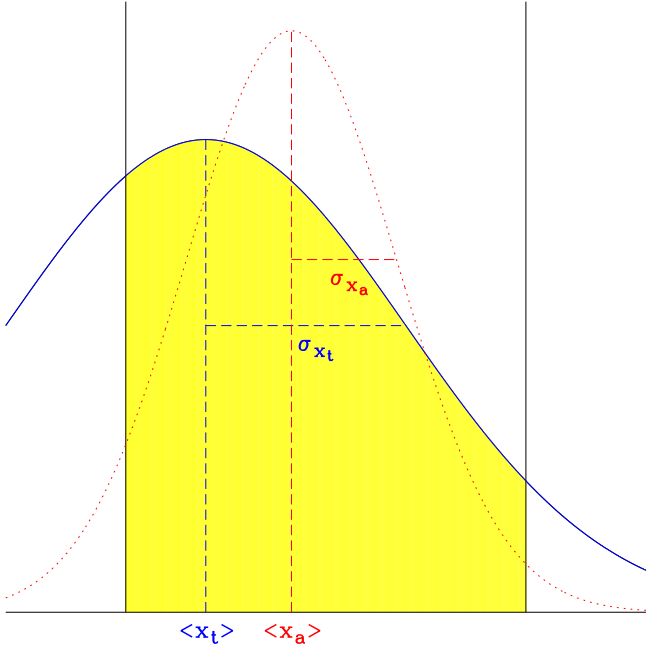}
\caption[]{Illustration of the change of the mean values and standard 
deviations of a Gaussian PDF by the limitation of the parameter range. The 
cut probability distribution (filled area) indicates a mean closer to the 
centre of the parameter range and an error smaller than the true value.}
\label{fig_illcorrppd}
\end{figure}

\begin{figure}
\centering 
\includegraphics[width=8.8cm,clip=true]{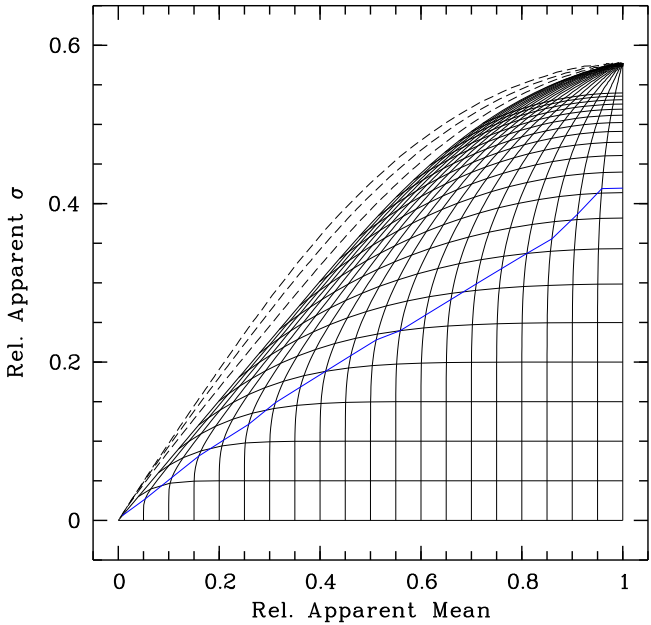}
\caption[]{Deviation of the measured from the real mean values and errors for
a Gaussian PDF. The axes show the apparent expectation values and standard 
deviations scaled to the half parameter range. The grid of solid curves 
indicates the true (and also scaled) mean values and errors for steps of 
$0.05$ from 0 to 1. Curves for constant real mean values $-0.3$, $-1$, and 
$-3$ outside the covered parameter range and variable real $\sigma$ are shown 
by dashed lines. The solid curve crossing the grid delimits the trustworthy 
area where the deviation between apparent and true mean values and errors is 
lower than 1\% and 10\%, respectively.}
\label{fig_corrppdcurv}
\end{figure}

\begin{figure}
\centering 
\includegraphics[width=8.8cm,clip=true]{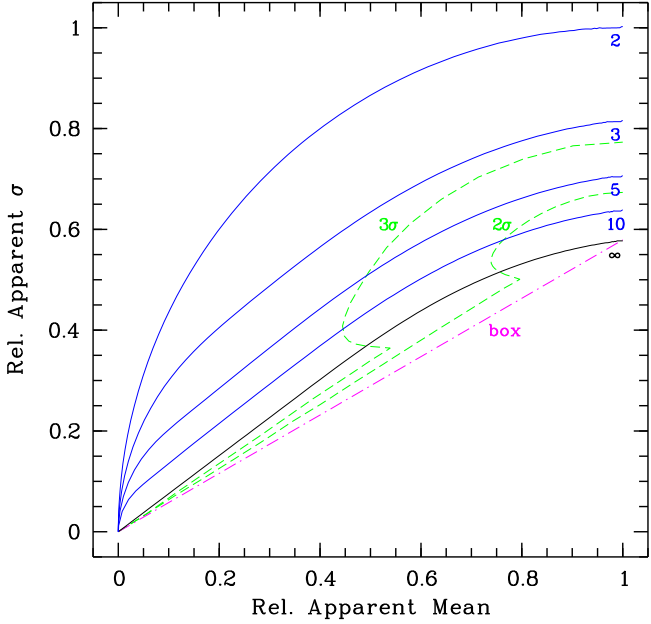}
\caption[]{Curves enveloping the area allowed for data in standardised 
diagrams showing mean values and errors scaled to the half parameter range.
The lowest solid curve corresponds to the enveloping curve of the grid shown
in Fig.~\ref{fig_corrppdcurv}, i.e., this curve is for a Gaussian PDF with 
the true mean inside the parameter range and an infinite number of parameter 
values. The dash-dotted curve shows the same for a box-shaped distribution 
instead of a Gaussian. The results for double Gaussians with $2\,\sigma$ and 
$3\,\sigma$ separation are marked by the dashed lines. The change in the 
data-covered area by the reduction of the parameter input values is indicated 
by the solid lines labelled by the number of values.}
\label{fig_envcurvdis}
\end{figure}

\begin{figure}
\centering 
\includegraphics[width=8.8cm,clip=true]{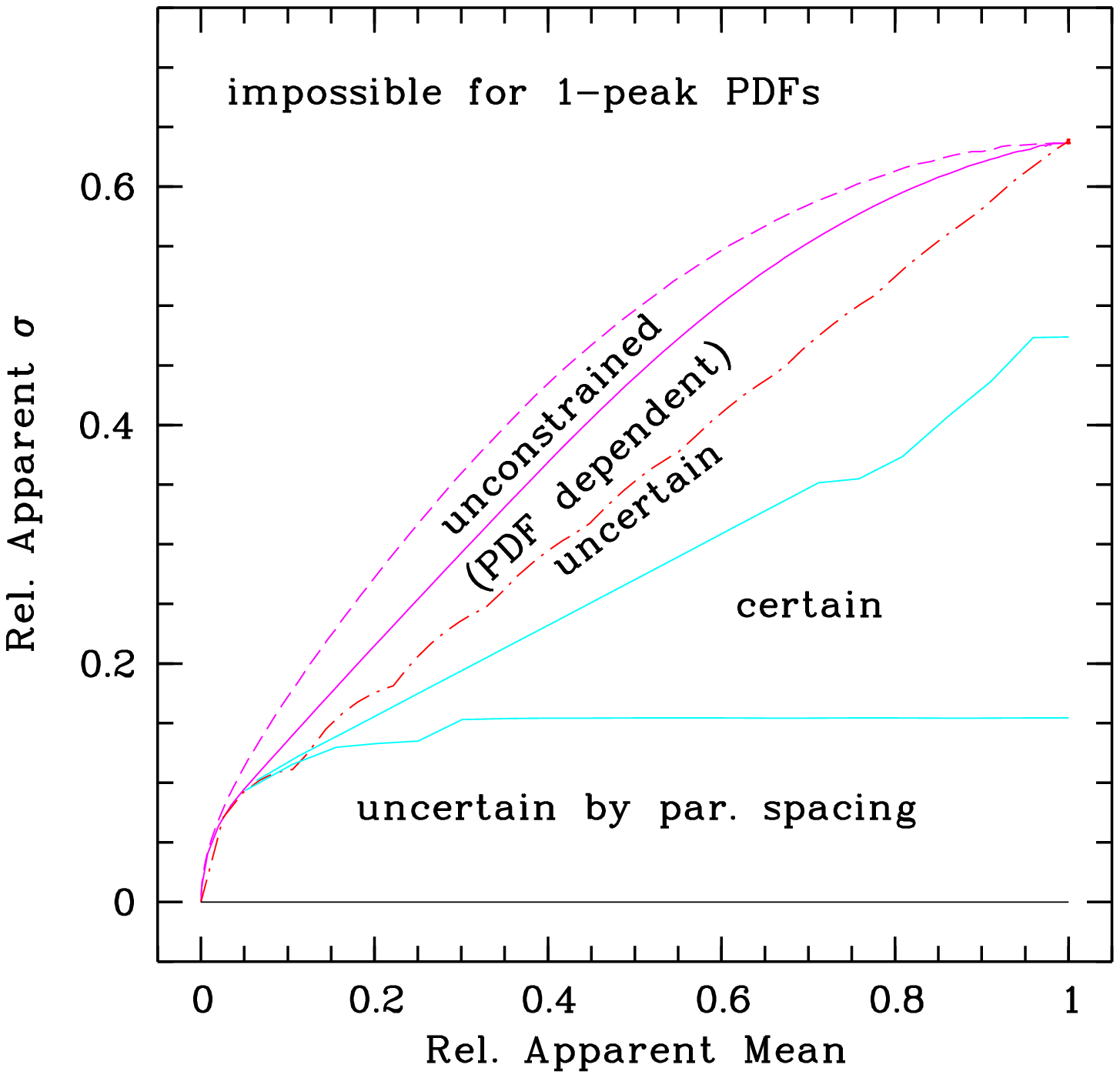}
\caption[]{Normalised diagnostic plot for the interpretation of mean values 
and standard deviations calculated by {\em CIGALE} for 10 equally-spaced 
parameter values between 0 and 2, i.e. a step size of $0.22$. The outer 
solid/dashed curves represent the enveloping curves for a Gaussian PDF having 
the mean inside/outside the parameter range. The dash-dotted curve marks the 
enveloping curve for a box-shaped distribution. The triangular area marked by 
``certain'' indicates for a Gaussian PDF the region of negligible difference 
between apparent and real mean values ($< 1\%$ of the half parameter range) 
and errors ($< 10\%$). For a box-shaped PDF this area would extend to higher 
$\sigma$ and be delimited by the corresponding enveloping curve. For applying 
the diagram to real data, the right part of the parameter range has to be 
complemented by the mirrored diagnostic curves and the axes have to be 
multiplied by half the difference of the maximum and minimum parameter values 
selected.}
\label{fig_illppddiag}
\end{figure}

The mean values $\langle x \rangle$ and errors $\sigma_x$ derived by 
{\em CIGALE} have to be taken with care. The high complexity of the models
and the limited number of photometric filters can cause that part of the
parameters are not well constrained. Moreover, restrictions of computing time 
and hard-disc space limit and/or inherent limitations of the models used 
restrict the range and number of possible values for each parameter which 
can be investigated by the code. Since good a-priori estimates of the PDF
of a parameter are rare, the selected model grid may fail to cover the entire 
set of parameter values consistent with the data. In this case the calculated 
average is probably misplaced and the errors are too small (see 
Fig.~\ref{fig_illcorrppd}). Therefore, we study the uncertainty and 
reliability of the results of {\em CIGALE} by means of simulations for 
different PDFs for different sets of parameter values. 

At first, we consider a Gaussian PDF for an infinite number of values. By 
changing the position and width of the Gaussian in a fixed parameter range of 
normalised width 2, we obtain the relation between apparent and true mean 
values and errors as presented in Fig.~\ref{fig_corrppdcurv}. For the area 
below the intersecting solid curve the regular grid in steps of $0.05$ does 
not change significantly, i.e. $\Delta \langle x \rangle < 0.01$ and 
$\Delta \sigma_x < 0.1$. For the centre of the Gaussian closer to the margin 
of the covered range and higher widths of the Gaussian the differences 
between true and apparent expectation values and standard deviations increase 
more and more until a curve that delimits the permitted area is reached. 
Close to this enveloping curve the corrections are enormous and highly 
uncertain because of the high density of curves. We can therefore divide the 
area of a standardised diagnostic diagram in three sub-areas which 
characterise reliable, uncertain/unreliable, and impossible results. 

For data points that are not too close to the enveloping curve, the diagram 
could, in principle, be used to recover the true mean values and errors. 
However, the real situation is more complicated, since PDFs do not need to be 
Gaussian (see, e.g., Figs.~\ref{fig_illppd_SFR} and \ref{fig_illppd_A_V}). 
Fig.~\ref{fig_envcurvdis} shows enveloping curves (for the true mean located 
inside the parameter range) for different shapes of the probability 
distribution. For a box-shaped distribution the possible errors are generally 
lower. On the other hand, double-peak profiles can cause relatively high 
apparent errors close to the centre of the parameter range. In this case, the 
true mean can be located in the other half of the parameter range, which is 
not possible for the single-peak PDFs discussed before.     

Finally, Fig.~\ref{fig_envcurvdis} illustrates that a finite number of 
parameter input values causes an increase of the errors. In the most extreme 
case of two values, only the margins of the parameter range are covered by 
data points, which produces significantly higher standard deviations than for 
a smooth PDF. In contrast to uncertainties in the shape of the probability 
distribution that impede a detailed diagnostics and correction, changes in 
the mean values and errors by the reduction of parameter values can be 
recovered and considered for the interpretation of the code results. 
Additional uncorrectable deviations are only caused if the true error is 
smaller than the difference between adjacent parameter values.  

In summary, diagnostic plots with similar curves as presented in this section
can be used to evaluate whether mean values and errors resulting from 
{\em CIGALE} are reliable, uncertain, or unreliable (implying an unconstrained 
parameter). Fig.~\ref{fig_illppddiag} illustrates these different areas in a 
diagnostic plot similar to those used for the interpretation of the code 
results in Sect.~\ref{properties}. Only for data points in or close to the 
``certain'' area the real PDF of a parameter can be well described by 
expectation value and standard deviation provided by {\em CIGALE}, otherwise 
the shape of the PDF has to be studied in more detail or it has to be 
accepted that the parameter cannot be constrained for the photometric data 
and model set given.

\section{A test sample: SINGS}\label{sings}

In order to establish {\em CIGALE} as a tool for studying properties of nearby
and distant star-forming galaxies, it has to be tested by using a well-known
reference data set. In the context of the Spitzer Infrared Nearby Galaxy 
Survey (SINGS; Kennicutt et al. \cite{KEN03}) high-quality photometric data 
were obtained in the IR regime for a sample of representative nearby 
galaxies. Together with the available photometric data for the other 
wavelength ranges (Dale et al. \cite{DAL07}; Mu\~noz-Mateos et al. 
\cite{MUN09}) this data set allows us to investigate the properties of the 
code in detail. We describe the sample in Sect.~\ref{sample} and discuss the 
analysis of the data set by means of {\em CIGALE} in Sect.~\ref{analysis}. 
The results and a comparison to the literature are presented in 
Sect.~\ref{results}.

\subsection{The sample}\label{sample}

\begin{table*}
\caption[]{Properties of the SINGS test sample. Although we list the 
relatively conservative run~B results (see Sect.~\ref{fitquality}), the given 
standard deviations could underestimate the real uncertainties because of the 
limited model grid used, model inaccuracies, and possible unconsidered 
photometric errors.}
\label{tab_SINGS}
\centering
\begin{tabular}{l l r@{.}l r@{\ $\pm$\ }l r@{\ $\pm$\ }l r@{\ $\pm$\ }l 
r@{\ $\pm$\ }l r@{\ $\pm$\ }l r@{\ $\pm$\ }l}
\hline\hline
\noalign{\smallskip}
ID & Type$^\mathrm{a}$ & \multicolumn{2}{c}{Dist.$^\mathrm{a}$} & 
\multicolumn{2}{c}{$\log M_\mathrm{star}$} & 
\multicolumn{2}{c}{$\log \mathrm{SFR}$} & 
\multicolumn{2}{c}{$\log t_{\,\mathrm{D4000}}$$^\mathrm{b}$} &
\multicolumn{2}{c}{$\log L_\mathrm{bol}$} & 
\multicolumn{2}{c}{$\log L_\mathrm{dust}$} &
\multicolumn{2}{c}{$A_\mathrm{FUV}$} \\
& & \multicolumn{2}{c}{[Mpc]} & 
\multicolumn{2}{c}{[M$_{\odot}$]} & 
\multicolumn{2}{c}{[M$_{\odot}$/yr]} & 
\multicolumn{2}{c}{[Gyr]} &
\multicolumn{2}{c}{[L$_{\odot}$]} & 
\multicolumn{2}{c}{[L$_{\odot}$]} & 
\multicolumn{2}{c}{[mag]} \\
\noalign{\smallskip}
\hline
\noalign{\smallskip}
NGC\,0024 & SAc    &  8 & 2 &  9.65 & 0.07 & -0.93 & 0.14 & -0.09 & 0.23 & 
 9.47 & 0.02 &  8.66 & 0.05 & 0.59 & 0.17 \\
NGC\,0584 & E4     & 27 & 6 & 11.39 & 0.03 & -1.10 & 0.33 &  0.93 & 0.07 & 
10.92 & 0.01 &  8.98 & 0.15 & 1.41 & 0.67 \\
NGC\,0925 & SABd   & 10 & 1 & 10.15 & 0.12 &  0.07 & 0.15 & -0.42 & 0.09 & 
10.26 & 0.03 &  9.68 & 0.04 & 0.70 & 0.15 \\
NGC\,1097 & SBb    & 16 & 9 & 11.35 & 0.10 &  0.84 & 0.17 &  0.00 & 0.26 & 
11.17 & 0.02 & 10.77 & 0.04 & 1.83 & 0.40 \\
NGC\,1291 & SBa    &  9 & 7 & 11.16 & 0.03 & -0.66 & 0.27 &  0.85 & 0.09 & 
10.70 & 0.01 &  9.37 & 0.07 & 1.21 & 0.59 \\
NGC\,1316 & SAB0   & 26 & 3 & 12.03 & 0.02 & -0.14 & 0.45 &  0.90 & 0.08 & 
11.56 & 0.01 & 10.10 & 0.06 & 1.59 & 0.86 \\
NGC\,1404 & E1     & 25 & 1 & 11.52 & 0.02 & -1.01 & 0.22 &  0.97 & 0.02 & 
11.04 & 0.01 &  8.46 & 0.10 & 0.48 & 0.16 \\
NGC\,1512 & SBab   & 10 & 4 & 10.43 & 0.08 & -0.28 & 0.20 &  0.15 & 0.22 & 
10.15 & 0.02 &  9.44 & 0.05 & 0.91 & 0.30 \\
NGC\,1566 & SABbc  & 18 & 0 & 11.05 & 0.13 &  0.88 & 0.18 & -0.35 & 0.08 & 
11.05 & 0.02 & 10.62 & 0.04 & 1.30 & 0.28 \\
NGC\,1705 & Am     &  5 & 8 &  8.51 & 0.08 & -1.48 & 0.19 & -0.52 & 0.03 &
 8.73 & 0.03 &  7.81 & 0.04 & 0.23 & 0.05 \\
NGC\,2798 & SBa    & 24 & 7 & 10.32 & 0.11 &  0.68 & 0.12 & -0.62 & 0.09 & 
10.64 & 0.02 & 10.51 & 0.03 & 4.53 & 0.32 \\
NGC\,2841 & SAb    &  9 & 8 & 10.99 & 0.03 & -0.31 & 0.24 &  0.60 & 0.18 & 
10.55 & 0.01 &  9.66 & 0.05 & 1.18 & 0.40 \\
NGC\,2976 & SAc    &  3 & 5 &  9.48 & 0.11 & -0.78 & 0.16 & -0.33 & 0.07 & 
 9.39 & 0.02 &  8.92 & 0.05 & 1.45 & 0.18 \\
NGC\,3031 & SAab   &  3 & 5 & 11.00 & 0.04 & -0.22 & 0.15 &  0.51 & 0.23 & 
10.60 & 0.01 &  9.60 & 0.05 & 0.90 & 0.36 \\
NGC\,3184 & SABcd  &  8 & 6 & 10.28 & 0.13 &  0.11 & 0.15 & -0.36 & 0.06 & 
10.26 & 0.01 &  9.76 & 0.04 & 1.13 & 0.19 \\
NGC\,3190 & SAap   & 17 & 4 & 10.92 & 0.03 & -0.85 & 0.51 &  0.82 & 0.12 & 
10.46 & 0.01 &  9.67 & 0.05 & 2.06 & 0.87 \\
NGC\,3198 & SBc    &  9 & 8 & 10.10 & 0.12 & -0.03 & 0.15 & -0.34 & 0.04 & 
10.13 & 0.01 &  9.60 & 0.04 & 0.89 & 0.17 \\
NGC\,3351 & SBb    &  9 & 3 & 10.66 & 0.08 & -0.04 & 0.19 &  0.17 & 0.18 & 
10.39 & 0.01 &  9.83 & 0.05 & 1.31 & 0.36 \\
NGC\,3521 & SABbc  &  9 & 0 & 11.00 & 0.04 &  0.43 & 0.22 & -0.28 & 0.03 & 
10.80 & 0.01 & 10.37 & 0.04 & 2.74 & 0.37 \\
NGC\,3621 & SAd    &  6 & 2 & 10.21 & 0.14 &  0.12 & 0.15 & -0.40 & 0.09 & 
10.28 & 0.02 &  9.88 & 0.04 & 1.31 & 0.26 \\
NGC\,3627 & SABb   &  8 & 9 & 10.95 & 0.05 &  0.55 & 0.22 & -0.29 & 0.03 &
10.79 & 0.02 & 10.40 & 0.05 & 2.31 & 0.31 \\
NGC\,4536 & SABbc  & 25 & 0 & 11.00 & 0.14 &  0.96 & 0.15 & -0.41 & 0.07 & 
11.12 & 0.02 & 10.82 & 0.04 & 2.00 & 0.33 \\
NGC\,4559 & SABcd  & 11 & 6 & 10.27 & 0.08 &  0.12 & 0.08 & -0.51 & 0.01 & 
10.45 & 0.01 &  9.90 & 0.04 & 0.80 & 0.18 \\
NGC\,4569 & SABab  & 20 & 0 & 11.41 & 0.02 &  0.49 & 0.22 &  0.33 & 0.08 & 
11.05 & 0.01 & 10.29 & 0.07 & 2.23 & 0.37 \\
NGC\,4579 & SABb   & 20 & 0 & 11.44 & 0.03 &  0.23 & 0.24 &  0.44 & 0.20 & 
11.04 & 0.01 & 10.21 & 0.05 & 2.02 & 0.57 \\
NGC\,4594 & SAa    & 13 & 7 & 11.75 & 0.02 & -0.47 & 0.47 &  0.92 & 0.05 & 
11.27 & 0.01 &  9.79 & 0.06 & 1.60 & 0.89 \\
NGC\,4625 & SABmp  &  9 & 5 &  9.35 & 0.13 & -0.72 & 0.11 & -0.34 & 0.05 & 
 9.35 & 0.02 &  8.83 & 0.04 & 0.86 & 0.19 \\
NGC\,4631 & SBd    &  9 & 0 & 10.47 & 0.14 &  0.70 & 0.20 & -0.62 & 0.08 & 
10.76 & 0.02 & 10.50 & 0.04 & 1.49 & 0.21 \\ 
NGC\,4725 & SABab  & 17 & 1 & 11.40 & 0.04 &  0.35 & 0.18 &  0.43 & 0.18 & 
11.01 & 0.00 & 10.14 & 0.04 & 1.12 & 0.33 \\
NGC\,4736 & SAab   &  5 & 3 & 10.80 & 0.07 &  0.07 & 0.20 &  0.11 & 0.23 & 
10.52 & 0.00 &  9.89 & 0.04 & 1.54 & 0.36 \\
NGC\,4826 & SAab   &  5 & 6 & 10.77 & 0.02 & -0.32 & 0.21 &  0.31 & 0.10 & 
10.40 & 0.00 &  9.60 & 0.04 & 2.38 & 0.34 \\
NGC\,5033 & SAc    & 13 & 3 & 10.77 & 0.11 &  0.31 & 0.20 & -0.27 & 0.06 & 
10.67 & 0.01 & 10.30 & 0.04 & 1.88 & 0.29 \\
NGC\,5055 & SAbc   &  8 & 2 & 10.96 & 0.06 &  0.43 & 0.22 & -0.28 & 0.02 & 
10.78 & 0.01 & 10.33 & 0.04 & 2.22 & 0.31 \\
NGC\,5194 & SABbc  &  8 & 2 & 10.82 & 0.12 &  0.79 & 0.15 & -0.44 & 0.07 & 
10.97 & 0.02 & 10.62 & 0.05 & 1.78 & 0.29 \\
NGC\,5195 & SB0p   &  8 & 2 & 10.77 & 0.02 & -0.12 & 0.21 &  0.33 & 0.08 & 
10.41 & 0.01 &  9.72 & 0.05 & 3.36 & 0.38 \\
NGC\,5474 & SAcd   &  6 & 9 &  9.30 & 0.06 & -0.65 & 0.14 & -0.45 & 0.04 & 
 9.50 & 0.02 &  8.59 & 0.05 & 0.27 & 0.06 \\
NGC\,5713 & SABbcp & 26 & 6 & 10.77 & 0.08 &  0.73 & 0.19 & -0.53 & 0.04 & 
10.93 & 0.01 & 10.71 & 0.03 & 2.86 & 0.33 \\
NGC\,5866 & S0     & 12 & 5 & 10.88 & 0.03 & -0.92 & 0.54 &  0.78 & 0.12 & 
10.43 & 0.00 &  9.30 & 0.06 & 2.51 & 1.18 \\
NGC\,7331 & SAb    & 15 & 7 & 11.39 & 0.08 &  0.78 & 0.20 & -0.04 & 0.25 & 
11.17 & 0.02 & 10.76 & 0.04 & 2.66 & 0.50 \\
\noalign{\smallskip}
\hline
\end{tabular}
\begin{list}{}{}
\item[$^\mathrm{a}$] see Kennicutt et al. (\cite{KEN03})
\item[$^\mathrm{b}$] Significant deviations of the metallicity from the 
solar value may affect the results for $t_{\,\mathrm{D4000}}$ (see 
Sect.~\ref{modelgrid}).
\end{list}
\end{table*}

\begin{figure}
\centering 
\includegraphics[width=8.8cm,clip=true]{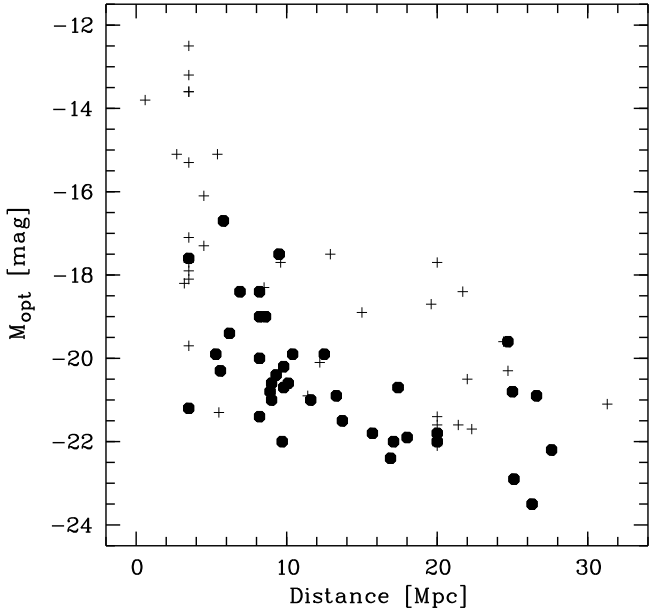}
\caption[]{Optical (i.e. mainly $R$) absolute magnitudes and distances in Mpc 
for the SINGS galaxies (see Kennicutt et al. \cite{KEN03}). The analysed
objects are indicated by filled symbols.}
\label{fig_mopt_dist}
\end{figure}

The full SINGS sample consists of 75 representative nearby galaxies 
(Kennicutt et al. \cite{KEN03}), i.e., the different morphological classes 
from ellipticals to irregulars have similar weights. Moreover, the galaxies
are distributed over roughly equal logarithmic bins in IR luminosity. Dale et 
al. (\cite{DAL07}, \cite{DAL08}) published UV-to-radio broad-bands SEDs for
the SINGS galaxies. The listed data comprises the $FUV$ ($\sim 1500$\,\AA{}) 
and $NUV$ ($\sim 2300$\,\AA{}) filters of {\em GALEX} (Gil de Paz et al. 
\cite{GIL07}), 2MASS data for $J$, $H$, and $K_\mathrm{s}$ (Jarrett et al. 
\cite{JAR03}), and IRAC and MIPS data for the filters centred on $3.6$, 
$4.5$, $5.8$, $8.0$, $24$, $70$, and $160$\,$\mu$m (Dale et al. 
\cite{DAL05}). Moreover, Dale et al. provide previously unpublished optical 
photometry for $B$, $V$, $R$, and $I$. However, since the optical data are 
erroneous\footnote{The corrected photometry provided in Dale et al. 
(\cite{DAL08}) is better than the original one given in Dale et al. 
(\cite{DAL07}). However, the results are still unsatisfying.}, we take the 
improved photometry of Mu\~noz-Mateos et al. (\cite{MUN09}), which is based 
on asymptotic magnitudes that were derived from surface photometry by 
extrapolating the curve of growth to encompass the whole galaxy light. Apart 
from all non-optical Dale et al. filters up to MIPS $160$\,$\mu$m, their 
catalogue provides photometry of the Sloan Digital Sky Survey (SDSS; 
Stoughton et al. \cite{STO02}) in $u'$, $g'$, $r'$, $i'$, and $z'$ for 32 
SINGS objects. Moreover, the Dale et al. (\cite{DAL07}, \cite{DAL08}) optical 
magnitudes of 23 galaxies without SDSS photometry were recalibrated by means 
of the catalogue of Prugniel \& Heraudeau (\cite{PRU98}), i.e., their 
aperture photometry was replicated on the SINGS optical images to derive 
zero-point corrections for each filter and each galaxy. For our sample 
selection we have required either the presence of SDSS photometry or at least 
three corrected Dale et al. filters. The latter criterion fulfil 16 of 23 
objects with recovered photometry only. Three of them do not have $R$-band 
photometry. Finally, we only take galaxies for which all UV and IR filters 
are available\footnote{Despite of the low quality of the far-IR data, two
elliptical galaxies were selected. NGC\,0584 was only marginally detected at 
160\,$\mu$m. Its flux at this wavelength could be contaminated by diffuse 
background/foreground emission. The fluxes at 70 and 160\,$\mu$m of NGC\,1404 
could be affected by a possible background source, which could not be masked 
and cleaned in a satisfying way due to the low image resolution at these 
wavelengths.}. Our selection criteria aim at ensuring a comprehensive 
photometric sample of similar high quality and wavelength coverage. 
Considering all criteria, our final sample comprises 39 SINGS galaxies (see
Table~\ref{tab_SINGS}). Fig.~\ref{fig_mopt_dist} shows the basic properties 
distance and absolute magnitude (mainly $R$) as given by Kennicutt et al. 
(\cite{KEN03}) for our subsample compared to the full sample. The diagram 
indicates that dwarf galaxies are underrepresented in our sample. There are 
no optical magnitudes for these galaxies in the catalogue of Mu\~noz-Mateos 
et al. (\cite{MUN09}) because of recalibration problems.

\subsection{Analysis}\label{analysis}

In the following we discuss the model grid that was selected to analyse our
test sample of SINGS galaxies (Sect.~\ref{modelgrid}). Moreover, the quality
of the fitting (Sect.~\ref{fitquality}) and the influence of the filter set 
on the results (Sect.~\ref{filterset}) is studied.

\subsubsection{The model grid}\label{modelgrid}

\begin{table}
\caption[]{Selected model parameter values for the analysis of our SINGS
sample}     
\label{tab_modpar}
\centering
\begin{tabular}{l c c c}
\hline\hline
\noalign{\smallskip}
Par. & Min. & Max. & $N$ \\
\noalign{\smallskip}
\hline
\noalign{\smallskip}
$Z$ [Z$_\odot$] & $1.0$ & $1.0$ & 1 \\
$\tau_\mathrm{oSP}$ [Gyr] & $0.25$ & $10.0$ & 4 \\
$t_{\,\mathrm{oSP}}$ [Gyr] & $10.0$ & $10.0$ & 1 \\
$\tau_\mathrm{ySP}$ [Gyr] & $0.05$ & $1.0$ & 3 \\
$t_{\,\mathrm{ySP}}$ [Gyr] & $0.05$ & $0.2$ & 2 \\
$f_\mathrm{burst}$ & $0.0001$ & $0.9999$ & 9 \\
$E_\mathrm{bump}$ & $0.0$ & $0.0$ & 1 \\ 
$c_\mathrm{R}$ & $0.0$ & $0.0$ & 1 \\
$\delta$ & $-0.3$ & $0.3$ & 5 \\
$A_{V,\,\mathrm{ySP}}$ & $0.0$ & $2.25$ & 16 \\ 
$f_\mathrm{att}$ & $0.0$ & $1.0$ & 5 \\
$\alpha$ & $1.0$ & $3.0$ & 9 \\
$f_\mathrm{AGN}$ & $0.0$ & $0.0$ & 1 \\ 
\noalign{\smallskip}
\hline
\end{tabular}
\end{table}

\begin{figure}[!ht]
\centering 
\includegraphics[width=7.65cm,clip=true]{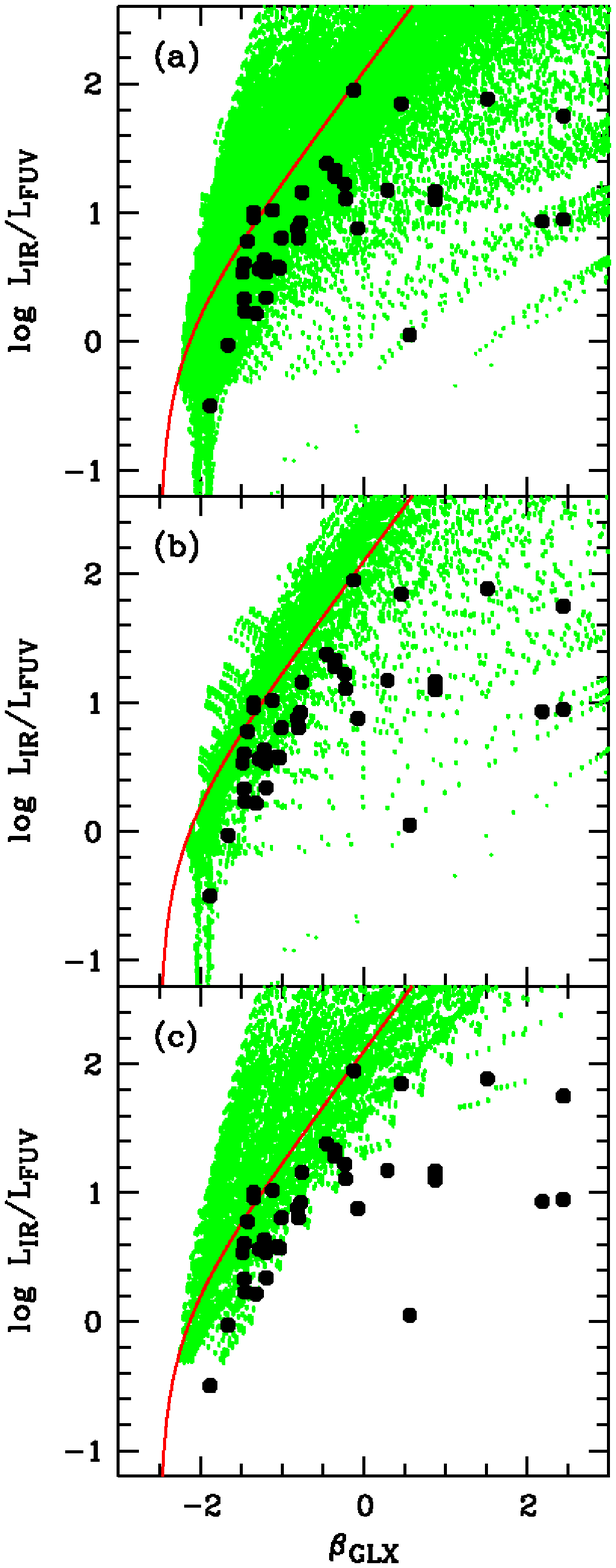}
\caption[]{The IR-to-far-UV luminosity ratio $L_\mathrm{IR}/L_\mathrm{FUV}$ 
versus the UV continuum slope $\beta_\mathrm{GLX}$ for the SINGS sample. The 
solid curve indicates the theoretical position of starbursts in the diagram 
(Kong et al. \cite{KON04}). The area in the 
$\beta_\mathrm{GLX}$--$L_\mathrm{IR}/L_\mathrm{FUV}$ plane that is covered by 
{\em CIGALE} models is marked by small dots. Subfigure (a) shows the full 
model set for the parameters given in Table~\ref{tab_modpar}. In (b) no slope 
correction of the Calzetti law is considered, i.e. $\delta = 0$. Finally, in 
(c) the stellar radiation is reddened regardless of the stellar population, 
i.e. $f_\mathrm{att} = 1$.}
\label{fig_kong}
\end{figure}

Studying the properties of our sample of 39 SINGS galaxies by SED fitting 
requires a careful selection of model parameters, since the parameter space 
has to be restricted for computing time reasons. Table~\ref{tab_modpar} 
shows the model grid that we have eventually used. The individual values were 
selected to provide (roughly) equally-sized steps in the given units or in 
dex for dynamical ranges larger than one order of magnitude. Only eight 
parameters have multiple input values, which is distinctly lower than the 15 
to 17 filters available for the SINGS galaxies in our sample. The total 
number of models is about $7 \times 10^5$, which can be managed quite easily 
on a typical PC because of a runtime of {\em CIGALE} of a few hours only. 

For all stellar population models we use Maraston (\cite{MARA05}) SEDs with 
Salpeter IMF and solar metallicity. The latter is justified by a mean 
oxygen abundance $\langle 12 + \log(\mathrm{O/H}) \rangle = 8.69$ and a 
scatter of $0.16$ for 23 sample galaxies listed in Calzetti et al. 
(\cite{CAL07}). Moreover, our sample lacks the low-metallicity dwarf galaxies 
present in the complete SINGS sample (see Fig.~\ref{fig_mopt_dist}). However, 
metallicity measurements are relatively uncertain due to high-quality 
requirements of the observations, insufficient coverage of a galaxy in 
combination with abundance gradients, and, in particular, calibration 
problems and general limitations of the metallicity tracer (see Moustakas \& 
Kennicutt \cite{MOU06} and references therein). Hence, we have checked the 
influence of the use of half and double solar metallicities on the results of 
the analysis of our SINGS sample. In general, the resulting properties 
indicate deviations within the errors and the corresponding uncertainties 
increase by 0 to 20\% only if metallicity uncertainties are considered. The 
only exception is the effective age measured at 4000\,\AA{} 
$t_{\,\mathrm{D4000}}$, which indicates a decrease of about $0.4$\,dex for a 
metallicity increase of a factor of 4. Consequently, the absolute values of 
$t_{\,\mathrm{D4000}}$ have to be taken with care due to the well-known 
age-metallicity degeneracy (e.g., Kodama \& Arimoto \cite{KOD97}).
   
In {\em CIGALE} SFHs are modelled by the combination of two models with 
exponentially decreasing SFR (see Sect.~\ref{stars}). For the ``old'' stellar 
population model we take a fixed age of 10\,Gyr, which should be a good 
compromise between the age of the Universe and the most important phase of 
star formation for the galaxies investigated. The $\tau$ values cover a 
relatively large range of values in order to account for the different kinds 
of SFHs possible. Due to the low effect of old stellar populations on galaxy 
SEDs, it is sufficient to select a few values only. For the ``young'' stellar 
population model we also take a few ages and $\tau$ only, since the variety 
of SEDs dominated by young stars is not very large. On the other hand, the 
mass fractions of both $\tau$ models are critical. Therefore, we analyse nine 
burst fractions $f_\mathrm{burst}$ covering the full range of theoretically 
possible values.
     
The obscuration of the stellar populations by dust is considered by the 
application of attenuation laws with different reasonable slopes (i.e. slopes
not too far from the Calzetti et al. (\cite{CAL00}) law) but without a UV 
bump. Since the UV wavelength range of the SINGS galaxies is covered by the 
two {\em GALEX} filters only, details of the attenuation law are difficult to 
study and insignificant for most galaxy properties as test runs of 
{\em CIGALE} indicate for a wide range of attenuation-related parameters. 
Therefore, we can restrict the number of models by only taking the default 
value zero of the $R_V$-related parameter $c_\mathrm{R}$ and by only 
considering slope corrections $\delta$ between $-0.3$ and $0.3$, which result 
in real $R_V$ between $3.0$ and $5.9$ (see Sect.~\ref{attenuation} and 
Fig.~\ref{fig_illattcurv}). Moreover, we fix the UV bump strength and select 
a value of zero. Apart from the negligible effect of the UV bump on the fit 
results for a wide range of strengths (cf. Noll et al. \cite{NOL09}), this 
choice is justified by the non-detection of a UV bump in local starburst 
galaxies (Calzetti et al. \cite{CAL94}), the lack of a significant 
2175\,\AA{} feature in the Kinney et al. (\cite{KIN96}) characteristic UV 
spectra of nearby galaxies covering a large range of star formation activity, 
the only low-to-moderate UV bump strengths in different galaxy populations 
(also comprising ``normal'' star-forming galaxies) studied at 
intermediate/high redshift (Noll et al. \cite{NOL09}), and the predicted 
weakening of the UV bump in local spiral galaxies by radiative transfer 
effects (Silva et al. \cite{SIL98}; Pierini et al. \cite{PIE04}).

Since the $V$-band attenuation of the young $\tau$ model 
$A_{V,\,\mathrm{ySP}}$ has an important effect on the (UV) SED slope and the 
UV-to-IR flux ratio, we choose a fine grid of values and a sufficiently high 
upper limit. The attenuation factor $f_\mathrm{att}$ indicates the fraction 
of $A_{V,\,\mathrm{ySP}}$ valid for the old $\tau$ model. The possibility to 
vary this parameter is crucial for fitting non-starburst galaxies as 
typically found in our SINGS sample. In Fig.~\ref{fig_kong} we show the 
luminosity ratio $L_\mathrm{IR}/L_\mathrm{FUV}$ and the UV continuum slope 
$\beta_\mathrm{GLX}$ of our sample galaxies. $L_\mathrm{IR}$ was derived from 
the MIPS 24, 70, and 160\,$\mu$m filters as described in Dale \& Helou 
(\cite{DAL02}). And based on {\em GALEX} $FUV$ and {\em GALEX} $FUV - NUV$,
respectively, $L_\mathrm{FUV}$ and $\beta_\mathrm{GLX}$ were calculated 
following the recipes of Kong et al. (\cite{KON04}). The objects are located 
close to or below the so-called ``Meurer'' relation (Meurer et al. 
\cite{MEU99}; Kong et al. \cite{KON04}). This curve is valid for starburst 
galaxies where the stellar content is completely obscured by a dust screen or 
shell. As the results of {\em CIGALE} in subfigure (c) imply, the galaxy 
distribution found cannot be explained by a deviation of the effective 
attenuation curve from the Calzetti law. Instead, subfigure (b) illustrates 
that an age-dependent amount of attenuation (cf. Silva et al. \cite{SIL98}; 
Pierini et al. \cite{PIE04}; Panuzzo et al. \cite{PAN07}; Noll et al. 
\cite{NOL07}) as simulated by $f_\mathrm{att}$ is able to reproduce the 
distribution of SINGS galaxies. Then, the data points that deviate most from 
the starburst curve are characterised by high obscuration of a young stellar 
component of very low mass (low $f_\mathrm{burst}$) and nearly unattenuated 
old stellar populations. Consequently, our code is suitable to study 
``normal'' star-forming galaxies which are usually located below the Meurer 
curve and are typical of the nearby Universe (see Buat et al. \cite{BUA05}; 
Cortese et al. \cite{COR06}; Dale et al. \cite{DAL07}; Salim et al. 
\cite{SALI07}).     

Dale \& Helou (\cite{DAL02}) offer IR dust emission templates in the 
$\alpha$ range from $0.0625$ to $4.0$. Our selection covers $\alpha$ values
between $1.0$ and $3.0$. The narrower range is reasonable, since 
$\alpha \ge 1.0$ correspond to mean IR luminosities 
$L_\mathrm{IR} \la 4 \times 10^{13}$\,L$_\odot$ according to the calibrations 
of Chapman et al. (\cite{CHAP03}) and Marcillac et al. (\cite{MARC06}). This 
is a safe limit even if the scatter in the calibrations is considered, since
the SINGS sample is characterised by relatively low IR luminosities (see 
Kennicutt et al. \cite{KEN03}). For very high $\alpha$ values the IR SEDs 
become very similar for the wavelength range studied. The calibrations of 
Chapman et al. and Marcillac et al., which are based on the ratio of the 
rest-frame fluxes at 60 and 100\,$\mu$m, are degenerated in this case. 
Therefore, it does not make sense to analyse $\alpha$ values beyond $3.0$.    

Finally, we do not use the option of an AGN contamination or hot-dust 
contribution for the main run. Although there are several Seyfert galaxies in 
the sample (Kennicutt et al. \cite{KEN03}), we do not expect that the 
{\em total} magnitudes of the selected galaxies are significantly affected by 
such a contribution. Nevertheless, we have checked with a different set of 
models allowing for a large range of hot-dust IR contributions whether 
{\em CIGALE} reproduces the expected result. Indeed, the vast majority of IR 
SEDs is consistent with no contribution at all. Only for the early-type galaxy 
NGC\,1404 we find a significant ($> 10$\%) hot-dust fraction. However, 
this result could also be explained by possible difficulties to fit the 
mixture of stellar radiation and very weak dust emission in the mid-IR of 
this special galaxy. Moreover, the fluxes at 70 and 160\,$\mu$m have to be 
taken with care due to a possible contamination by a background source (see
note in Table~\ref{tab_SINGS}).

\subsubsection{The fit quality}\label{fitquality}

\begin{table}
\caption[]{Mean photometric errors for our SINGS test sample}     
\label{tab_photerr}
\centering
\begin{tabular}{l c}
\hline\hline
\noalign{\smallskip}
Filters & Rel. errors \\
\noalign{\smallskip}
\hline
\noalign{\smallskip}
GALEX $FUV$, $NUV$ & 15\% \\
Dale et al. $B$, $V$, $R$, $I$ & 16\% \\ 
SDSS $u'$, $g'$, $r'$, $i'$, $z'$ & 3\% \\
2MASS $J$, $H$, $K_\mathrm{s}$ & 1\% \\
IRAC $3.6$, $4.5$, $5.8$, $8.0$\,$\mu$m & 11\% \\
MIPS $24$\,$\mu$m & 5\% \\
MIPS $70$\,$\mu$m & 7\% \\
MIPS $160$\,$\mu$m & 13\% \\
\noalign{\smallskip}
\hline
\end{tabular}
\end{table}

\begin{figure}
\centering 
\includegraphics[width=8.8cm,clip=true]{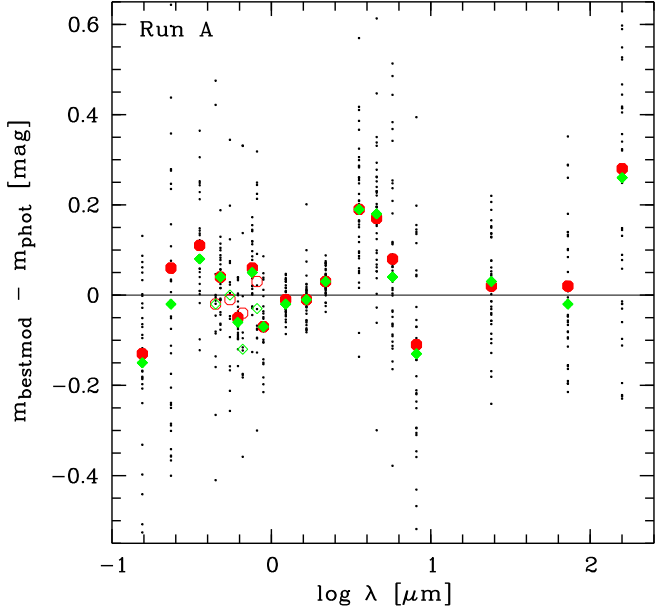}
\caption[]{Ratio of the best-fit model fluxes and the measured object fluxes 
in mag for run~A and different filters and objects. The filters are 
indicated by their mean wavelength. Sample-related mean and median deviations 
for each filter are marked by big circles and lozenges, respectively. The 
open symbols indicate the recalibrated optical filters of Dale et al. 
(\cite{DAL07}).}
\label{fig_filtdev}
\end{figure}

\begin{figure}
\centering 
\includegraphics[width=7.7cm,clip=true]{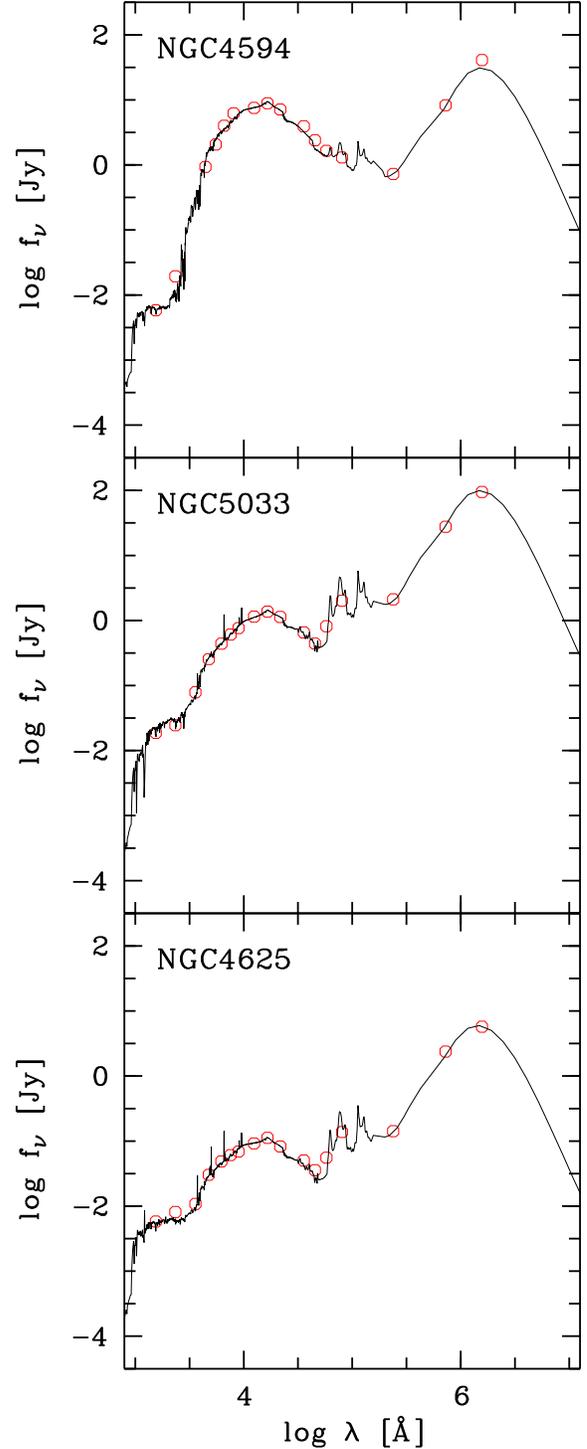}
\caption[]{Illustration of three characteristic best-fit SEDs (run~A) which 
differ in their UV-to-optical flux ratios. From top to bottom, the SEDs of 
NGC\,4594 (Sa), NGC\,5033 (Sc), and NGC\,4625 (Sm) are shown. The measured
photometric fluxes are marked by circles.}
\label{fig_plotbestsed}
\end{figure}

For the analysis of the properties of our SINGS test sample we use the total
magnitudes of 21 filters and its errors reported in Mu\~noz-Mateos et al. 
(\cite{MUN09}). The typical uncertainties for the different filters are 
listed in Table~\ref{tab_photerr}. They are dominated by the 
instrument-related zero-point errors or uncertainties of photometric 
corrections and vary between 1\% for the 2MASS near-IR filters and 16\% for 
the Dale et al. (\cite{DAL07}) optical magnitudes corrected by Mu\~noz-Mateos 
et al. (\cite{MUN09}). The latter are only relevant for 15 of 39 sample 
galaxies for which SDSS photometry of much higher precision is not available. 
Since the relative photometric errors represent the weights for the different 
filters in the fitting process, the weights of the optical magnitudes for the 
two subsamples with and without SDSS photometry are quite different. 
Therefore, we will discuss the results of both subsamples separately in 
Sect.~\ref{results}. 

The large differences in the reported uncertainties for the filter fluxes 
obtained by different projects and instruments raises the question whether 
these values are really comparable. The homogeneity of the error evaluation 
is crucial for reliable fitting results. In particular, very low errors such 
as reported for the 2MASS filters are critical, since this causes hard 
constraints for the model set. Moreover, especially small errors are 
questionable, since possible systematic offsets between different filters can 
cause erroneous fits. Since the models are not a perfect reproduction of the 
nature, systematic errors are also expected for them if the fluxes of 
different wavelength ranges are compared. For most wavelengths we assume 
uncertainties in the order of 5 to 10\%. Higher values are especially 
expected for the rest-frame wavelength range between $2.5$ and 5\,$\mu$m for 
which the continuum is relatively uncertain due to the mixing of stellar and 
dust emission and fundamental uncertainties in the shape of both components. 
An example is the insufficient knowledge of the contribution of TP-AGB stars 
at these wavelengths (see Maraston \cite{MARA05}). Therefore, we study two 
different photometric catalogues. In the first case (`run~A') we take the 
published errors (see Table~\ref{tab_photerr}) and in the second case 
(`run~B') we increase them by adding an additional moderate 5\% error in 
quadrature to all filter errors. This procedure allows us to investigate how 
the results change if additional systematic uncertainties are considered. 

As a first result of {\em CIGALE}, Fig.~\ref{fig_filtdev} shows the deviations 
of the best-fit models from the object photometry in magnitudes for each
filter. Since the best-fit models of both runs indicate similar photometric 
fluxes, we only plot the data of run~A. In general, the sample-averaged 
deviations for the different filters are relatively small. On average, the 
difference is $0.07$\,mag only. The worst filters are MIPS 160\,$\mu$m, IRAC 
$3.6$\,$\mu$m, and IRAC $4.5$\,$\mu$m with deviations amounting to $0.28$, 
$0.19$, and $0.17$\,mag. The modest fit quality for these three filters is 
caused by relatively high uncertainties in the object photometry and the 
shape of the models in the corresponding wavelength ranges (see above). For 
MIPS 160\,$\mu$m the significant deviations can partly be explained by the 
lacking flexibility of the one-parameter models of Dale \& Helou 
(\cite{DAL02}) in reproducing the complex IR SEDs of real galaxies. On the 
other hand, the differences between the Dale \& Helou templates of adjacent 
$\alpha$ values (see Table~\ref{tab_modpar}) are distinctly larger for MIPS 
160\,$\mu$m than for any other IR filter studied. The corresponding 
uncertainties are similar to those indicated in Fig.~\ref{fig_filtdev}. This 
error source becomes negligible if expectation values derived from the 
parameter PDFs are analysed instead of best-fit models.
   
The good fit quality that we can reach with the parameter set chosen over a
wide wavelength range is demonstrated in Fig.~\ref{fig_plotbestsed}, which
shows the best-fit SEDs of three SINGS galaxies in comparison to the measured
photometry. Although the dust reddening and star formation activity of 
NGC\,4594, NGC\,5033, and NGC\,4625 differ considerably, the photometric SEDs
of all galaxies shown are reproduced quite well.  
                  
Finally, we have checked whether our approach to derive PDFs has any 
influence on the results. In general, we find that the differences between 
the `max' and `sum' methods (see Sect.~\ref{fitting}) are negligible, i.e., 
the differences in the expectation values are distinctly smaller than the 
errors. For example, the `sum' method (for a lower model confidence limit of 
$10^{-3}$) changes the sample-averaged total stellar mass and SFR by only 
$-0.01$\,dex ($\sigma = 0.03$) and $+0.04$\,dex ($\sigma = 0.09$) for run~A, 
respectively. In comparison, the corresponding errors are $0.05$\,dex and 
$0.16$\,dex. They increase by 12\% and 2\%, respectively, by the use of the 
`sum' method instead of our preferred `max' method. Although the 
sample-averaged differences are small, significant effects cannot be excluded 
for individual galaxies, however. The most extreme deviation for the SFR is 
an increase of $0.34$\,dex for NGC\,5866. It is caused by diminishing, 
additional peaks in the probability distribution at low SFRs due to a low 
number of models populating these peaks. Although such cases are rare as the 
comparison of the galaxy properties derived by both approaches 
shows\footnote{The differences in the SFRs of the `max' and `sum' methods are 
relatively large in comparison to the results for other model parameters.}, 
it is the reason why we prefer the `max' method, which reduces the influence 
of the model density on the weighing of parameter values.

\subsubsection{Influence of filter set on results}\label{filterset}

\begin{figure}
\centering 
\includegraphics[width=8.8cm,clip=true]{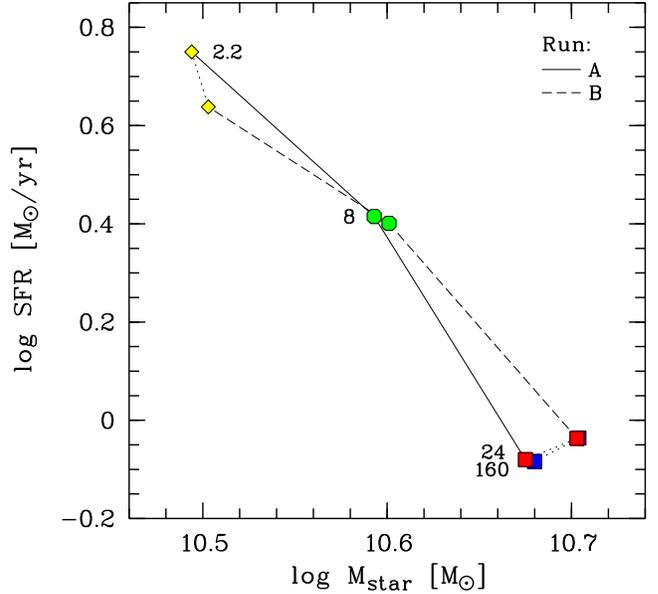}
\caption[]{Effect of the reduction of IR filters on the basic galaxy 
properties total stellar mass and SFR. The diagram provides the sample means 
and mean errors of these quantities for run~A (solid lines) and run~B (dashed 
lines) and different filter combinations. The labels near the symbols 
indicate the mean wavelength of the last filter in the filter set in $\mu$m. 
$160$ means that all filters from GALEX $FUV$ to MIPS $160$\,$\mu$m are used 
for the analysis. $2.2$ refers to the $K_\mathrm{s}$ band as the reddest 
filter.}
\label{fig_SFR_mass}
\end{figure}

\begin{table}
\caption[]{Change of the SINGS sample mean for different properties by 
running {\em CIGALE} without IR data}     
\label{tab_efffilt}
\centering
\begin{tabular}{l c c}
\hline\hline
\noalign{\smallskip}
Par. & Run~A & Run~B \\
\noalign{\smallskip}
\hline
\noalign{\smallskip}
$\log M_\mathrm{star}$ & $-0.19$ & $-0.20$ \\
$\log \mathrm{SFR}$ & $+0.83$ & $+0.67$ \\
$\log L_\mathrm{bol}$ & $+0.37$ & $+0.31$ \\ 
$\log L_\mathrm{dust}$ & $+0.75$ & $+0.15$ \\
$\log f_\mathrm{burst}$ & $+1.05$ & $+0.72$ \\
$\log t_{\,\mathrm{D4000}}$ & $-0.78$ & $-0.71$ \\
$A_\mathrm{FUV}$ & $+2.16$ & $+1.59$ \\
$A_V$ & $+0.65$ & $+0.57$ \\ 
$\alpha$ & $-0.40$ & $-0.44$ \\
 \noalign{\smallskip}
\hline
\end{tabular}
\end{table}

Since {\em CIGALE} allows us to fit photometric data of galaxies ranging from 
far-UV to far-IR, star-formation-related parameters can be derived in a 
reliable way due to a full coverage of the dust-affected wavelength range.
However, for distant galaxies especially far-IR data are often not available 
or not sensitive enough. Therefore, we have investigated how results would 
change if the IR filter set of our sample galaxies was incomplete.    

Fig.~\ref{fig_SFR_mass} shows the sample-averaged total stellar mass and SFR
for a reduction of the IR filter set. While the full filter set contains 
160\,$\mu$m as the reddest filter, the most reduced filter set ends at 
$2.2$\,$\mu$m, i.e. the $K_\mathrm{s}$ band. There is no significant change 
of the results as long as the filter set still includes at least one filter 
(i.e. 24\,$\mu$m) which traces the dust emission beyond the PAH regime. 
Amplitude and slope of the Dale \& Helou (\cite{DAL02}) models appear to be 
well determined in this case. In contrast, the sample means change 
considerably if the IR is only covered up to 8\,$\mu$m, i.e. the region of 
the PAH emission features, where the Dale \& Helou models show little variety 
only. For run~A, $M_\mathrm{star}$ decreases by $0.19$\,dex and the SFR 
increases by $0.83$\,dex if $K_\mathrm{s}$ is the reddest filter. These 
numbers show that the disregard of mid-IR and far-IR data results in 
unsatisfying fits for our SINGS test sample. Table~\ref{tab_efffilt} shows 
that many parameters are concerned. Apart from the SFR, important changes are 
also present for the dust luminosity ($+0.75$\,dex), the burst fraction 
($+1.05$\,dex), and $t_{\,\mathrm{D4000}}$ ($-0.78$\,dex). The effective 
attenuation factors in the far-UV $A_\mathrm{FUV}$ and visual $A_V$ 
considerably increase by $2.16$\,mag and $0.65$\,mag, respectively (cf. 
Burgarella et al. \cite{BUR05}), if IR data are not considered. Compared to 
other properties the effect on the mass is relatively small, since this 
quantity is only indirectly affected by a higher but still minor importance 
of young stellar populations and the corresponding change of the 
mass-to-light ratio. 

Fig.~\ref{fig_SFR_mass} and Table~\ref{tab_efffilt} also illustrate the 
results for higher uncertainties in the filter fluxes (run~B), which indicate 
lower deviations for most properties. In particular, the changes for 
$L_\mathrm{dust}$ ($+0.15$\,dex), $f_\mathrm{burst}$ ($+0.72$\,dex), and 
$A_\mathrm{FUV}$ ($+1.59$\,mag) are significantly reduced. This suggests that 
the low photometric errors in some filters of run~A could force the code into 
unfavourable fits if no IR data is available. Consistently, the galaxies 
with SDSS data and consequently lower photometric errors in the optical 
indicate higher differences in the mean SFR of code runs with and without IR 
data than objects with corrected optical Dale et al. photometry only.

We have also studied the code results for a filter set without the GALEX $FUV$ 
and $NUV$ filters. The differences for most parameters are insignificant and 
the SFR increases by $0.17$\,dex (run~A) or $0.11$\,dex (run~B) only. Even
$A_\mathrm{FUV}$ that directly depends on the UV flux changes by $+0.02$\,mag 
for run~A only. An exception is the increase of this quantity by $0.46$\,mag 
for run~B. However, this amount is still much lower than the deviations found 
for the neglection of IR data. Consequently, the UV data of the SINGS 
galaxies have only little influence on the fit results. This is probably 
caused by the relatively low weight of these filters in the fitting 
process (see Table~\ref{tab_photerr}). The relatively high uncertainties in 
the photometry of these two filters do not appear to significantly constrain 
the star-formation-sensitive UV SEDs of the sample galaxies. Consequently, a 
lack of IR data can cause striking systematic deviations in the galaxy 
properties. However, the amount of these deviations probably depends on the 
SFHs of the galaxies investigated and the filter set available. For example, 
missing IR data could be better compensated at higher redshifts because of a 
better filter coverage in the rest-frame UV due to the shift of the observed 
frame of optical telescopes.

\subsection{Results}\label{results}

In the following we present the physical properties of our SINGS sample 
obtained by means of {\em CIGALE} and discuss their reliability 
(Sect.~\ref{properties}). We also show relations between the different 
parameters and compare our results to those from other studies 
(Sect.~\ref{correlations}).

\subsubsection{Expectation values and standard deviations}\label{properties}

\begin{figure*}
\centering 
\includegraphics[width=18cm,clip=true]{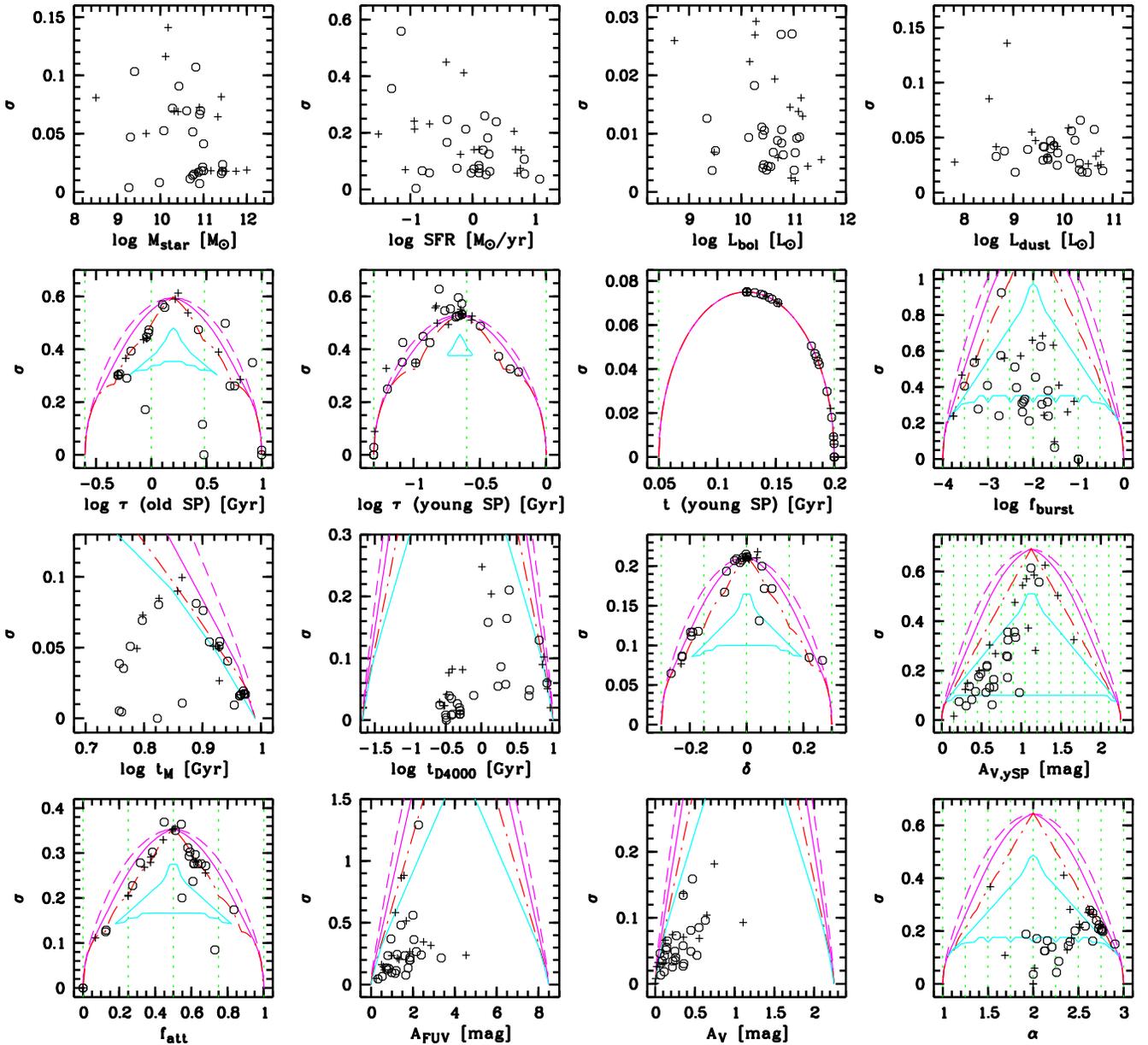}
\caption[]{Run~A code results for our SINGS test samples with and without 
SDSS photometry (circles and crosses, respectively). Expectation values and 
standard deviations are shown for mass-dependent properties 
($M_\mathrm{star}$, SFR, $L_\mathrm{bol}$, and $L_\mathrm{dust}$), basic 
model parameters with at least two selected input values 
($\tau_\mathrm{oSP}$, $\tau_\mathrm{ySP}$, $t_{\,\mathrm{ySP}}$, 
$f_\mathrm{burst}$, $\delta$, $A_{V,\,\mathrm{ySP}}$, $f_\mathrm{att}$, and 
$\alpha$), complex ages ($t_{\,\mathrm{M}}$ and $t_{\,\mathrm{D4000}}$), and 
effective dust attenuation factors ($A_\mathrm{FUV}$ and $A_V$). For the 
basic parameters the selected parameter values are indicated by vertical 
dotted lines. The different diagnostic curves are described in 
Fig.~\ref{fig_illppddiag}. All curves depend on the number of parameter 
values given. For the complex ages and effective attenuation factors we 
assume 50 which is the number of bins of the parameter probability 
distributions. The curves for the mass-dependent parameters are not shown, 
since the errors are much lower than the widths of the object-dependent 
parameter ranges.}
\label{fig_expandstda}
\end{figure*}

\begin{figure*}
\centering 
\includegraphics[width=18cm,clip=true]{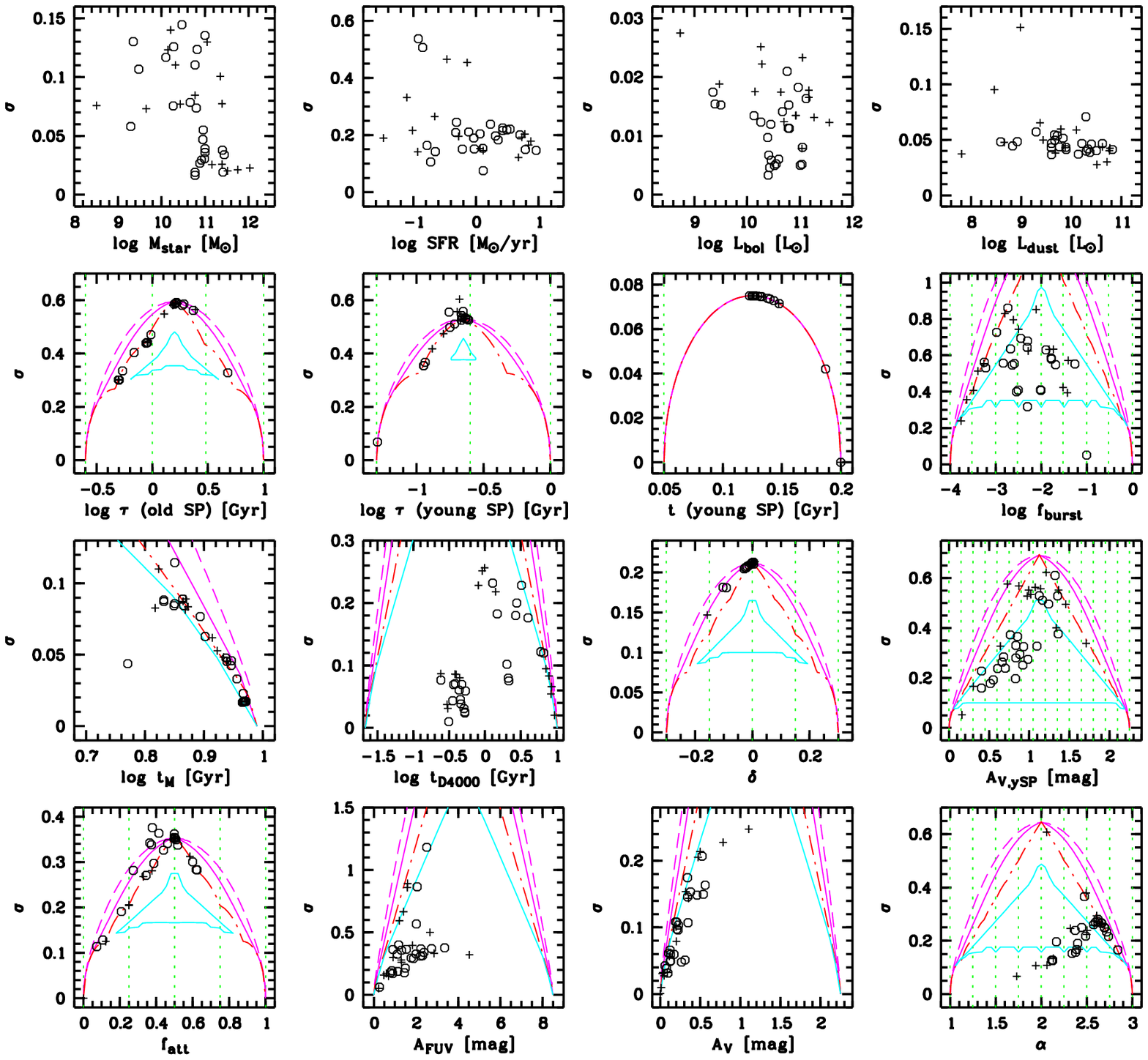}
\caption[]{Alternative run~B code results for our SINGS test samples. In 
comparison to Fig.~\ref{fig_expandstda}, the given errors of the input 
photometry were increased by 5\% added in quadrature in order to account for 
additional systematic photometric errors and model-inherent uncertainties.}
\label{fig_expandstdb}
\end{figure*}

In Fig.~\ref{fig_expandstda} (run~A) and Fig.~\ref{fig_expandstdb} (run~B) 
we illustrate the expectation values and standard deviations for the different 
model parameters of {\em CIGALE} for our test sample of 39 SINGS galaxies. We
only show results for parameters with PDFs consisting of at least two values. 
Data points based on photometry with and without SDSS photometry are marked 
by different symbols (circles and crosses). For evaluating the reliability of 
the results, the diagrams with mass-independent properties also indicate the 
diagnostic curves introduced in Sect.~\ref{interpretation}. In fact, the 
enveloping curves for Gaussian and box-shaped PDFs as well as the ``safe'' 
areas for Gaussian distributions are shown.

At first, we discuss the mass-dependent properties $M_\mathrm{star}$, SFR, 
$L_\mathrm{bol}$, and $L_\mathrm{dust}$ (see Table~\ref{tab_SINGS}). Typical 
total stellar masses of about $10^{11}$\,M$_{\odot}$ and SFRs between $0.1$ 
and 10\,M$_{\odot}$yr$^{-1}$ reflect that the sample mainly consists of 
nearby spiral galaxies (see Sect.~\ref{sample}). Moreover, the dust 
luminosities are lower than $10^{11}$\,L$_{\odot}$, which indicates that 
luminous IR galaxies (LIRGs) are not present in our sample. The 
mass-dependent properties are quite well constrained (see 
Fig.~\ref{fig_illppd_SFR} for an example PDF). The typical run~A errors range 
from $0.011$\,dex for $L_\mathrm{bol}$ to $0.16$\,dex for the SFR, which is 
distinctly smaller than the range of parameter values covered by the sample. 
This statement is also true for the somewhat higher average errors between 
$0.013$ and $0.22$\,dex for the less constrained run~B. The differences 
between the parameter values for run~A and run~B are usually within the 
errors. For the most uncertain mass-dependent property, the SFR, the 
sample-averaged value increases by $0.05$\,dex only if the SFRs of run~B 
instead of run~A are considered (see Fig.~\ref{fig_SFR_mass}). The mean 
difference between the SFRs of both runs for individual galaxies is 
$0.10$\,dex.     

\begin{figure}
\centering 
\includegraphics[width=8.8cm,clip=true]{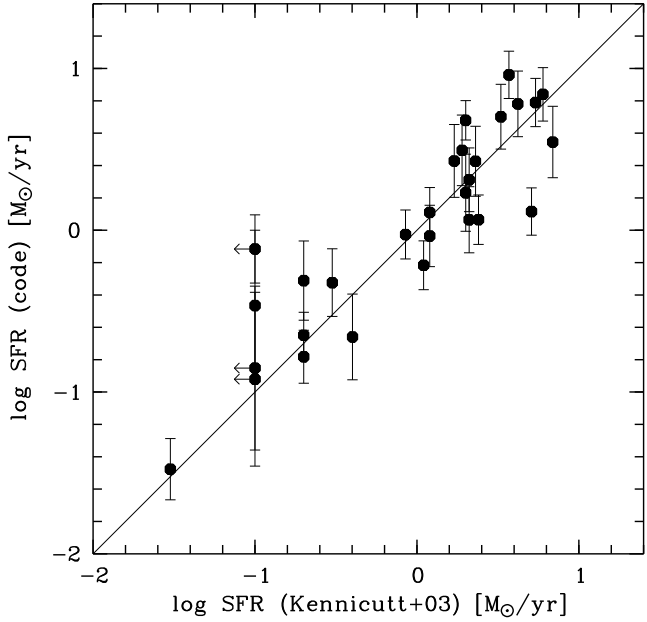}
\caption[]{Comparison of the SFRs derived from run~B of {\em CIGALE} for the 
SINGS sample and those of Kennicutt et al. (\cite{KEN03}) based on H$\alpha$ 
emission. Upper limits are indicated by arrows. The error bars of the code 
results indicate the standard deviation of the SFR probability distribution 
for each galaxy.}
\label{fig_compSFR}
\end{figure}

In contrast to basic model parameters which describe the SFH or the 
attenuation of the stellar radiation (see Table~\ref{tab_pardesc}), the 
mass-dependent global galaxy properties are relatively independent of the 
details of the models used. Hence, the results for our SINGS test sample can 
be compared quite comfortably to other studies in this 
case\footnote{Unfortunately, the completely different parameter set and a 
significantly different selection of test galaxies from SINGS impedes a 
qualitative and quantitative comparison of our code results to those of da 
Cunha et al. (\cite{CUN08}).}. The dust luminosity $L_\mathrm{dust}$ is 
relatively easy to measure, since it is directly related to the filter fluxes 
in the IR (provided that the filters cover the entire relevant wavelength 
range). Nevertheless, the comparison of $L_\mathrm{dust}$ obtained by 
different studies is a good consistency check. For 13 sample galaxies 
$L_\mathrm{dust}$ determinations are also available by Draine et al. 
(\cite{DRA07b}) who derived them from the SINGS IR SEDs using the Draine \& 
Li (\cite{DRA07a}) dust models. The mean difference between our estimates 
and those of Draine et al. is negligible, since it results in 
$0.00 \pm 0.03$~dex for run~A and B. The scatter amounts to $0.1$\,dex. A 
comparison of instantaneous SFRs is more crucial, in particular, if the 
approaches are completely different. For 29 sample galaxies SFRs of 
Kennicutt et al. (\cite{KEN03}) are available. They are based on H$\alpha$ 
emission, which is a thoroughly studied and relatively reliable SFR indicator 
(see, e.g., Kennicutt \cite{KEN98}; Brinchmann et al. \cite{BRI04}; Kewley et 
al. \cite{KEW04}; Calzetti et al. \cite{CAL07}). Despite of the completely 
different methods, our SFRs of run~B exhibit a moderate relative scatter of 
$0.3$\,dex and deviate on average by $0.06 \pm 0.05$~dex only from those of 
Kennicutt et al. (see Fig.~\ref{fig_compSFR}). For run~A the SFRs agree even 
better, since the mean difference amounts to $0.00 \pm 0.06$~dex. In any 
case, this comparison shows that our results for our SINGS test sample are 
reliable at least for galaxy properties which are relatively independent of 
the details of the modelling. 

The basic stellar-population-related parameters are the ages and $\tau$ of
two Maraston (\cite{MARA05}) models, which are linked by the burst fraction 
$f_\mathrm{burst}$ (see Sect.~\ref{stars} and Table~\ref{tab_modpar}). We 
show them in Figs.~\ref{fig_expandstda} and \ref{fig_expandstdb} excepting 
the constant $t_{\,\mathrm{oSP}}$. For $\tau_\mathrm{oSP}$ and run~A the 
resulting values cover almost the entire investigated range between $0.25$ 
and $10$\,Gyr. However, the errors are relatively high. Most of the data 
points are close to or on the enveloping curves. In particular, the 
distribution appears to follow the enveloping curve for box-shaped PDFs. If 
those distributions were typical (which is difficult to prove because of the 
small number of parameter values), the $\tau_\mathrm{oSP}$ of most galaxies 
could be classified as ``rather low'' or ``rather high'' only. However, the 
situation appears to be even worse if the results of run~B are considered. 
The lower accuracy of the input photometry causes that for the majority of 
the objects $\tau_\mathrm{oSP}$ is not constrained at all, as the clustering 
of data points in the peak of the enveloping curves indicates. Consequently, 
it is safer to assume that the given photometry is not accurate enough to 
derive $\tau$ values for the old stellar population model. This also appears 
to be the case for the age and $\tau$ of the young stellar population model. 
There is only a trend towards relatively high mean $t_{\,\mathrm{ySP}}$ for 
run~A. However, this trend vanishes completely if the results of run~B are 
considered. In contrast, the burst fractions of most sample galaxies are 
reliable at least in the context of the SFHs allowed by the model grid. Even 
for run~B the data points cluster inside the area of trustworthy results. If 
run~A provided the correct results, the errors in $f_\mathrm{burst}$ of part 
of the galaxies would be even below the grid spacing, which is suggested by 
a position below the ``triangle'' in the diagnostic diagram. In any case, 
small burst fractions below 1\% dominate the distribution. No galaxy shows 
$f_\mathrm{burst} > 10\%$. This result implies that the current 
star formation activity tends to be lower than the past average, since the 
age ratio of the old and young stellar population model is about 80.
  
The relatively high degree of degeneracy in the SEDs for different SFHs and 
the enormous amount of possible evolutionary scenarios prevents the reliable 
derivation of most basic model parameters related to the stellar population. 
Hence, we also discuss the effective ages $t_{\,\mathrm{M}}$ and 
$t_{\,\mathrm{D4000}}$ (see Sect.~\ref{stars}). The mean mass-weighted age 
$t_{\,\mathrm{M}}$ of 8\,Gyr is close to the age of the old stellar 
population model of 10\,Gyr. Consequently, the galaxies under study appear to 
be dominated by old stars that were formed in the first half of the lifetime 
of the Universe. However, the exact $t_{\,\mathrm{M}}$ of the individual 
galaxies is relatively uncertain. While one third of the objects indicates a 
trustworthy age in the case of run~A, there are only a few galaxies with 
reliable ages in the case of run~B. The decrease in errors around 6\,Gyr is 
related to the maximum burst fraction of 10\% found in the sample. The age 
$t_{\,\mathrm{D4000}}$ derived from the strength of the 4000\,\AA{} break 
indicates a wider variety than $t_{\,\mathrm{M}}$ and is quite well 
constrained even for run~B (see also Table~\ref{tab_SINGS}) and even under 
consideration of the metallicity dependency of the results (see 
Sect.~\ref{modelgrid}). If the mean $t_{\,\mathrm{D4000}}$ of about 1\,Gyr 
is taken, the sample can be divided into a spectroscopically young subsample 
of 22 galaxies and a mean age of 400\,Myr and a complementary old one of 17 
galaxies with 3.6\,Gyr on average. The relatively wide range in 
$t_{\,\mathrm{D4000}}$ reflects the variety of galaxy types in SINGS ranging 
from ellipticals to irregulars (see Kennicutt et al. \cite{KEN03}). While 
90\% of the young subsample indicates morphological types of Sb and later, 
only 20\% of the old subsample exhibits these types.
 
Since we do not scale the attenuation law by a $R_V$-like constant factor and 
do not consider the possible presence of 2175\,\AA{} absorption features in 
the SEDs (see Sect.~\ref{modelgrid}), the shape of the attenuation curve is 
only determined by the deviation $\delta$ of the slope from a Calzetti law. 
Figs.~\ref{fig_expandstda} and \ref{fig_expandstdb} show that $\delta$ of 
many galaxies is not constrained at all and the rest is highly uncertain. 
Interestingly, significant deviations of $\delta$ from zero are almost 
exclusively found for run~A and galaxies with SDSS photometry. The latter has 
not played a role for the parameters discussed so far. This result implies 
that differences in the shape of the attenuation curve can only be detected 
if the accuracy of the photometry is in the order of a few percent. For the 
SINGS sample the accuracy is not good enough for any reliable statements. The 
situation would probably be better if more and better UV data were available 
for the SINGS galaxies. However, this is not the case. In comparison to 
$\delta$ the scaling parameter $A_{V,\,\mathrm{ySP}}$ for the young stellar 
population, which ranges from $0.15$ to $1.66$ for run~A, can be constrained 
quite well. Most galaxies are in the reliable region of the diagnostic 
diagram. However, the individual values significantly depend on the run, in 
particular, if SDSS photometry is available. In this case, the mean 
$A_{V,\,\mathrm{ySP}}$ changes from $0.67$ for run~A to $0.87$ for run~B. For 
the other galaxies the average increase is only half as strong (from $0.89$ 
to $1.00$). This behaviour can be explained by the stronger run~A SED 
constraints in the optical for galaxies with SDSS photometry and the coupling 
of $\delta$ and $A_{V,\,\mathrm{ySP}}$ that is suggested by the stability of 
the dust luminosity. The steeper slopes of the attenuation law for run~A 
($\langle \delta \rangle = -0.04$ versus $-0.01$ for run~B) have to be 
compensated by lower $A_{V,\,\mathrm{ySP}}$ in order to reach a similar 
obscuration of the stellar populations by dust. Another attenuation-related 
parameter is $f_\mathrm{att}$ which gives the fraction of 
$A_{V,\,\mathrm{ySP}}$ for the old stellar population. For individual 
galaxies the interpretation of the results is uncertain especially for run~B. 
On the other hand, the mean $f_\mathrm{att}$ for run~A and run~B are 
$0.45 \pm 0.03$ and $0.41 \pm 0.02$, respectively. This suggests a trend 
towards $f_\mathrm{att} < 0.5$ in SINGS. The result is consistent with the 
fact that many galaxies in our SINGS sample are highly diversified galaxies 
for which the amount of dust attenuation depends on the age of the stellar 
population (cf. Fig.~\ref{fig_kong}).

The results for the attenuation at 1500\,\AA{} are reliable for most 
galaxies. The parameter space allowed by the model grid is only partly 
covered by sample objects. Relatively low $A_\mathrm{FUV}$ dominate. The mean
values are $1.5$ (run~A) and $1.6$ (run~B). They correspond to 78\% and 69\%,
respectively, of the values expected for the sample $A_{V,\,\mathrm{ySP}}$ 
and the Calzetti law ($\delta = 0$), which is characterised by 
$A_\mathrm{FUV}/A_V = 2.55$. Hence, the old $\tau$ model ($t/\tau \sim 10$) 
significantly contributes even in the far-UV, which is obviously due to the 
very low burst fractions and low $f_\mathrm{att}$ of most sample galaxies. 
The effective attenuation factors in the visual $A_V$ are less reliable than 
$A_\mathrm{FUV}$ and $A_{V,\,\mathrm{ySP}}$. In particular, the run~B results 
are relatively uncertain. This outcome can be explained by the strong 
dependence of $A_V$ on the relatively uncertain attenuation factor 
$f_\mathrm{att}$. The mean $A_V$ of both runs is about $0.3$\,mag, which 
corresponds to mean $A_V/A_{V,\,\mathrm{ySP}}$ of $0.38$ (run~A) and $0.29$ 
(run~B). These values are comparable to the $f_\mathrm{att}$ of the sample 
and suggest that the optical continuum of most sample galaxies is almost 
completely given by the old $\tau$ model.       
  
\begin{figure*}
\centering 
\includegraphics[width=11.25cm,clip=true,angle=-90]{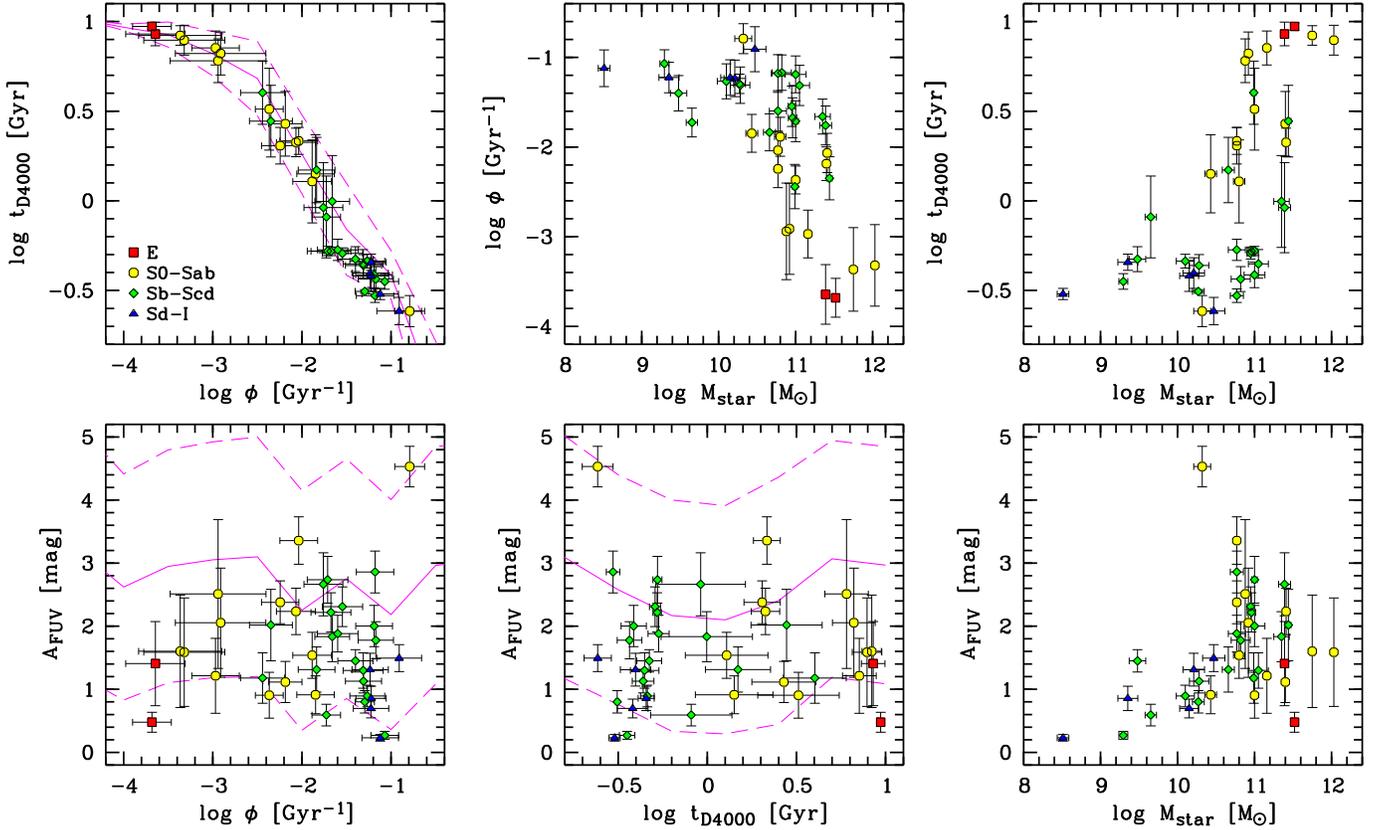}
\caption[]{Relations between the total stellar mass $M_\mathrm{star}$, the 
specific SFR $\phi$, the effective age at 4000\,\AA{} $t_{\,\mathrm{D4000}}$, 
and the attenuation in the far-UV $A_\mathrm{FUV}$ for the SINGS test sample. 
The given values and errors are related to run~B of {\em CIGALE}. The 
different symbols indicate different morphologies (see legend in upper left 
subfigure). The solid and dashed curves in the plots not related to 
$M_\mathrm{star}$ show the mean values and $1\,\sigma$ scatter of the model 
distribution.}
\label{fig_plotcorrel}
\end{figure*}

Finally, we discuss the shape of the dust emission in the IR. Since we do not 
consider a hot dust continuum (see Sect.~\ref{modelgrid}), it is only
determined by the $\alpha$ parameter of the Dale \& Helou (\cite{DAL02})
templates (see Sect.~\ref{dustemission}). Figs.~\ref{fig_expandstda} and 
\ref{fig_expandstdb} indicate that low $\alpha$ are very well constrained, 
while high $\alpha$ are relatively uncertain. This is obviously caused 
by the degeneracy of the dust emission models for wavelengths shortwards of 
the emission peak for high $\alpha$. For $\alpha > 2.5$ the flux ratio 
$f_{\nu}(60\,\mu\mathrm{m})/f_{\nu}(100\,\mu\mathrm{m})$ becomes almost 
constant. Relatively high $\alpha$ dominate the distribution, i.e., the dust 
temperature tends to be cool. The average sample value of $2.4$ can be 
compared to the mean $L_\mathrm{dust}$ of $6 \times 10^9$\,L$_{\odot}$ by 
using the calibrations of Chapman et al. (\cite{CHAP03}) and Marcillac et al. 
(\cite{MARC06}) for which we obtain $\alpha \approx 2.1$ and $2.3$. The 
moderate differences in $\alpha$ could be explained by different properties
of the samples used for the calibrations and the SINGS sample. Differences
could also be caused by the degeneracy in 
$f_{\nu}(60\,\mu\mathrm{m})/f_{\nu}(100\,\mu\mathrm{m})$ for high $\alpha$ 
values and the fact that this flux ratio was taken for the Chapman et al. and 
Marcillac et al. calibrations instead of $\alpha$.

\subsubsection{Correlations}\label{correlations}

The different galaxy properties derived by {\em CIGALE} are not completely
independent of each other. For example, the burst fraction $f_\mathrm{burst}$
and the two effective ages $t_{\,\mathrm{M}}$ and $t_{\,\mathrm{D4000}}$ are 
well correlated, since the stellar population properties of the models used 
mainly depend on the burst fraction (see Sect.~\ref{properties}). Another 
example is the relation between the slope modification of the Calzetti law 
$\delta$ and the attenuation in the visual of the young $\tau$ model 
$A_{V,\,\mathrm{ySP}}$ based on a relatively stable dust luminosity 
$L_\mathrm{dust}$ discussed in Sect.~\ref{properties}. Finally, 
$M_\mathrm{star}$, SFR, $L_\mathrm{bol}$, and $L_\mathrm{dust}$ tend to be 
correlated because of their dependence on the galaxy mass (see 
Table~\ref{tab_pardesc}). Therefore, we restrict our discussion of possible 
correlations between different galaxy properties on parameters for which the 
models do not show a direct relation. Moreover, we only consider parameters 
which were well determined in run~B of {\em CIGALE} for our SINGS sample (see 
Table~\ref{tab_SINGS}) and, therefore, do not suffer from crucial 
degeneracies. In fact, we discuss the {\em CIGALE} output parameters 
$M_\mathrm{star}$, $t_{\,\mathrm{D4000}}$, and $A_\mathrm{FUV}$. Moreover, we 
show the specific SFR $\phi$, i.e. the instantaneous SFR divided 
by the total mass produced by star formation in the past\footnote{For 
Maraston (\cite{MARA05}) models the time-integrated SFR corresponds to the 
galaxy mass, which comprises the total stellar mass and the mass of gas 
released from stars by winds/explosions (see Sect.~\ref{stars}).}.
  
Figure~\ref{fig_plotcorrel} presents the mutual relations of 
$M_\mathrm{star}$, $\phi$, $t_{\,\mathrm{D4000}}$, and $A_\mathrm{FUV}$ for 
our sample of 39 SINGS galaxies. The tightest correlation is found for the 
quantities $\phi$ and $t_{\,\mathrm{D4000}}$, which depend on stellar 
population parameters only. It shows that effectively younger stellar 
populations are linked to higher star formation activity. The correlation 
obtained is in good qualitative and quantitative agreement with the results 
of Brinchmann et al. (\cite{BRI04}) for the relation between H$\alpha$-based 
$\phi$ and D4000 of star-forming SDSS galaxies. However, our distribution is 
narrower, since a tight correlation is already inherent in the code and the 
model grid used as the distribution of models in the 
$\phi$--$t_{\,\mathrm{D4000}}$ plane indicates. For the 
$M_\mathrm{star}$-related diagrams in Fig.~\ref{fig_plotcorrel} correlations 
based on the selected model grid or the internal structure of the models are 
not possible, since the mass is the free scaling parameter in the fitting 
process. Therefore, all mass-related correlations found must have an 
astrophysical interpretation only. Although the correlation is weaker than 
those discussed before, the specific SFR clearly tends to decrease with 
increasing mass, which is in qualitative agreement with previous studies 
(see, e.g., Brinchmann et al. \cite{BRI04}; Salim et al. \cite{SALI07}; Buat 
et al. \cite{BUA07}, \cite{BUA08}). Quantitatively, there are some 
differences due to different sample properties. For example, Salim et al.
(\cite{SALI07}) find an average $\phi$ for their sample of star-forming SDSS
galaxies at $z \approx 0.1$ that is about $0.4$\,dex higher at 
$10^{10}$\,M$_{\odot}$ than the value of $-1.3$\,dex typical of our sample. 
The transition from active star formation to a relatively passive evolution 
is at about $10^{11}$\,M$_{\odot}$ for our SINGS galaxies. This break in the 
galaxy properties is even more pronounced for the effective age 
$t_{\,\mathrm{D4000}}$. The bimodality found is consistent with the results 
of Kauffmann et al. (\cite{KAU03a}, \cite{KAU03b}) for the SDSS and is 
reminiscent of the two basic galaxy populations ``red sequence'' ($\ge 7$ 
objects) and ``blue cloud'' ($\ge 19$ objects) discussed in, e.g., Strateva 
et al. (\cite{STR01}), Baldry et al. (\cite{BALD04}), and Driver et al. 
(\cite{DRI06}). The galaxies in-between have the highest age uncertainties 
and could represent at least in part the so-called ``green valley''.

In Fig.~\ref{fig_plotcorrel} we also identify the morphological types of our
sample galaxies as given in Table~\ref{tab_SINGS} (see Kennicutt et al. 
\cite{KEN03}). For the relations discussed so far, the locus of a galaxy is 
clearly related to the morphological properties. The earlier the type of a 
galaxy is, the lower $\phi$ and the higher $t_{\,\mathrm{D4000}}$ and 
$M_\mathrm{star}$ are. These correlations show that our code results are 
consistent with the well-known dependence of the star formation activity and 
mass of nearby galaxies on the morphological type (see Roberts \& Haynes 
\cite{ROB94} and Kennicutt \cite{KEN98} for reviews). In this context, the 
extreme properties of the early-type SBa spiral galaxy NGC\,2798 can be 
understood by the presence of a nuclear starburst that produces a high amount 
of deeply dust-enshrouded young stars.

\begin{figure}
\centering 
\includegraphics[width=8.8cm,clip=true]{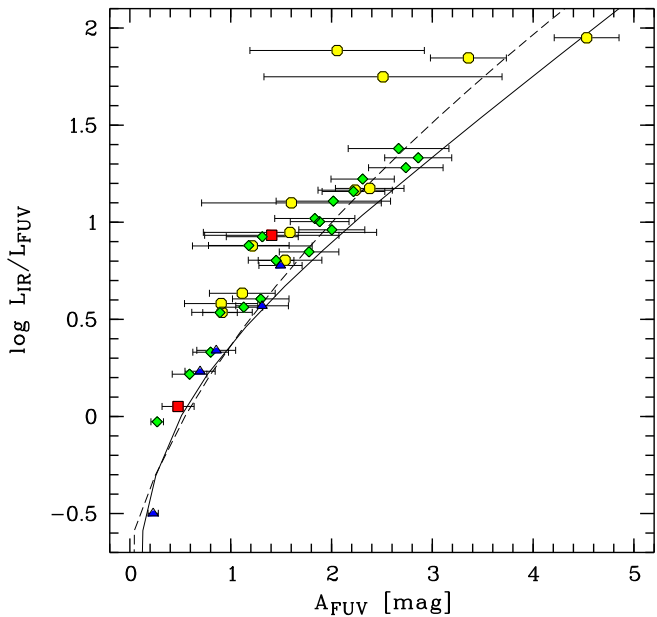}
\caption[]{Comparison of the attenuation in the far-UV $A_\mathrm{FUV}$ 
(run~B) and the luminosity ratio $L_\mathrm{IR}/L_\mathrm{FUV}$ as defined 
by Kong et al. (\cite{KON04}) for our SINGS test sample. For the meaning of 
the symbols see Fig.~\ref{fig_plotcorrel}. Errors can only be shown for the 
code results. The solid curve exhibits the relation between both parameters 
for a solar metallicity Maraston (\cite{MARA05}) model with constant SFR and 
an age of 100\,Myr that is reddened by a Calzetti law. The dashed curve shows 
the corresponding relation for nearby star-forming galaxies of Burgarella et 
al. (\cite{BUR05}).}
\label{fig_compatttracer}
\end{figure}

No convincing correlation is found for the attenuation at 1500\,\AA{} 
$A_\mathrm{FUV}$ and $\phi$ or $t_{\,\mathrm{D4000}}$, i.e. the obscuration 
of the young stellar population is relatively independent of the 
star formation activity and history. Concerning the morphology, the latest 
types (and possibly the ellipticals) in the sample tend to have lower 
$A_\mathrm{FUV}$ than most spiral galaxies. For instance, $A_\mathrm{FUV}$ 
equals $2.00 \pm 0.29$ for S0--Sab, $1.64 \pm 0.17$ for Sb--Scd, and 
$0.92 \pm 0.22$ for Sd--I. Nevertheless, this trend is much weaker than those 
found by Buat \& Xu (\cite{BUA96}) and Dale et al. (\cite{DAL07}) based on 
the attenuation tracer $L_\mathrm{IR}/L_\mathrm{FUV}$ (cf. 
Fig.~\ref{fig_kong}) instead of $A_\mathrm{FUV}$. The latter authors have 
also found an anticorrelation of $L_\mathrm{IR}/L_\mathrm{FUV}$ and the 
specific SFR for intermediate- to late-type galaxies, which cannot be 
confirmed by us. A possible explanation of these discrepancies could be a 
systematic weakness of $L_\mathrm{IR}/L_\mathrm{FUV}$. In contrast to 
$L_\mathrm{FUV}$, $L_\mathrm{IR}$ is related to the obscuration of young 
{\em and} old stars. Therefore, significant deviations should appear where 
dust-absorbed light from cool stars significantly contributes to 
$L_\mathrm{IR}$ (see, e.g., Buat et al. \cite{BUA05}; Cortese et al. 
\cite{COR08} and references therein). The distribution of the sample galaxies 
in the $A_\mathrm{FUV}$--$L_\mathrm{IR}/L_\mathrm{FUV}$ plane shown in 
Fig.~\ref{fig_compatttracer} in comparison to the locus of a typical 
starburst SED for different opacities indicates that early-type spirals 
probably exhibit the largest deviations, while the latest types are close to
the starburst curve. In detail, the mean deviations of 
$L_\mathrm{IR}/L_\mathrm{FUV}$ from the reference curve amount to 
$0.31 \pm 0.07$ for S0--Sab, $0.17 \pm 0.02$ for Sb--Scd, and $0.03 \pm 0.05$ 
for Sd--I. As indicated by Fig.~\ref{fig_compatttracer} early-type spirals 
also exhibit the largest deviations from the relation between 
$L_\mathrm{IR}/L_\mathrm{FUV}$ and $A_\mathrm{FUV}$ for nearby star-forming 
galaxies derived by Burgarella et al. (\cite{BUR05}) (cf. Buat et al. 
\cite{BUA05}; Cortese et al. \cite{COR08}). Consequently, the apparent trend 
of $L_\mathrm{IR}/L_\mathrm{FUV}$ with morphology and $\phi$ could be 
strengthened/produced by the SFH dependence of the luminosity ratio used.   

A reliable trend probably exists for the relation between $A_\mathrm{FUV}$ 
and $M_\mathrm{star}$. The highest obscurations appear to be present for 
masses around $10^{11}$\,M$_{\odot}$, i.e. for galaxies in the transition 
region. On the other hand, there is also a relatively large scatter in 
$A_\mathrm{FUV}$ of high mass galaxies. For low mass galaxies the dust 
obscuration tends to decrease with decreasing mass. Late-type galaxies of 
low mass are obviously characterised by lower dust column densities or a 
higher porosity of the dust screen in front of young stars compared to 
galaxies of earlier type (cf. Dale et al. \cite{DAL07}). Apart from 
irregular and screen-diluting dust distributions, a ``lack'' of UV-absorbing 
dust grains could also be due to a deficit of dust production compared to 
dust destruction or a low metallicity.

\section{Discussion and conclusions}\label{discussion}

We have developed the SED-fitting code {\em CIGALE} as a tool for studying 
basic properties of galaxies in the near and the distant Universe. The models
constructed by {\em CIGALE} consist of Maraston (\cite{MARA05}) or PEGASE
(Fioc \& Rocca-Volmerange \cite{FIO97}) stellar population models which are 
reddened by synthetic attenuation curves based on the Calzetti et al. 
(\cite{CAL00}) law and which are corrected for spectral lines, and the dust 
emission templates of Dale \& Helou (\cite{DAL02}). The construction of the 
SFH by two different complex stellar population models with exponentially 
declining SFRs and different amounts of attenuation enables the code to deal 
with age-dependent extinction effects (Silva et al. \cite{SIL98}; Kong et al. 
\cite{KON04}; Panuzzo et al. \cite{PAN07}; Noll et al. \cite{NOL07}), which 
is crucial to fit ``normal'' star-forming galaxies that are not characterised 
by a uniform dust screen in front of the stars (see Fig.~\ref{fig_kong}). The 
models cover the wavelength range from far-UV to far-IR, which allows the 
effect of dust on galaxy SEDs to be treated in a consistent way. In the case 
of dust emission related to a non-thermal source, the balance between the 
stellar luminosity absorbed by dust and the corresponding emitted luminosity 
in the IR can be preserved by considering an additional hot dust component.   

As our study of the multi-wavelength photometric data of a test sample of 39 
nearby galaxies selected from SINGS (Kennicutt et al. \cite{KEN03}; Dale et 
al. \cite{DAL07}; Mu\~noz-Mateos et al. \cite{MUN09}) shows, especially the 
star-formation-related properties can be derived with high reliability if
the photometry reaches from the rest-frame UV to wavelengths greater than
10\,$\mu$m (Sect.~\ref{filterset}). Otherwise, the multi-parameter models are 
not well constrained in the IR and the results can be affected by systematic 
errors. In the latter case, the models have to be simplified by reasonable
a-priori estimates of part of the parameters. Apart from the wavelength 
coverage of the filter set, the individual uncertainties in filter fluxes 
also play an important role for the quality of the fit results. Large 
uncertainties can prevent the derivation of parameters that are nearly 
degenerated regarding the shape of the galaxy SED such as details of the SFH 
or the attenuation curve. However, it is more critical if too low 
photometric errors overconstrain the model parameters, yielding relatively 
precise but possibly wrong results. Due to unknown systematic errors in the 
object photometry and the models (see Sect.~\ref{fitquality}) it is advisable 
to assume relative errors between 5\% and 10\% in minimum. In contrast, the 
selected approach for the derivation of the expectation values and standard 
deviations of the different parameters does not significantly affect the code 
results at least on average (Sect.~\ref{fitquality}). The `max' method 
introduced by us (Sect.~\ref{fitting}) and the `sum' method (e.g., Kauffmann 
et al. \cite{KAU03a}; Salim et al. \cite{SALI07}; Walcher et al. 
\cite{WAL08}) give similar results as long as there are no secondary peaks in 
the parameter probability distribution which are only poorly populated by 
models.     

The diagnostics of the SED-fitting results of our SINGS test sample has 
revealed that the most reliable values are obtained for non-basic (by the 
code derived) model properties such as the total stellar mass 
$M_\mathrm{star}$, the SFR, the effective age $t_{\,\mathrm{D4000}}$, the 
bolometric luminosity $L_\mathrm{bol}$, the dust luminosity 
$L_\mathrm{dust}$, and the far-UV attenuation factor $A_\mathrm{FUV}$ 
(Sect.~\ref{properties}). Trustworthy results are also found for basic input 
parameters of the models such as the burst fraction $f_\mathrm{burst}$, the 
$V$-band dust attenuation of the young stellar population 
$A_{V,\,\mathrm{ySP}}$, and in part the slope of the IR models $\alpha$. 
However, the latter properties significantly depend on the model grid chosen 
and are not universal, therefore. The non-basic properties usually show weak 
model-related constraints only. An example is the upper 
$t_{\,\mathrm{D4000}}$ limit of 10\,Gyr in the sample due to our restrictions 
regarding the SFHs investigated by the code (Sect.~\ref{modelgrid}). For the 
masses given, it has to be taken into account that we provide results for 
Maraston (\cite{MARA05}) models and Salpeter IMF (see Sect.~\ref{stars}). In 
any case, the SFR and $L_\mathrm{dust}$ derived indicate good agreement with 
data from other studies (Sect.~\ref{properties}).

An investigation of relations between the different reliable galaxy 
properties has confirmed that the star formation activity of nearby galaxies
as traced by the specific SFR $\phi$ and the effective age 
$t_{\,\mathrm{D4000}}$ depend on morphology (Sect.~\ref{correlations}). 
Weaker trends are also found for $M_\mathrm{star}$ and $A_\mathrm{FUV}$. The 
typical star formation activity significantly changes at 
$M_\mathrm{star} \sim 10^{11}$\,M$_{\odot}$. This mass range also indicates 
the most dust-obscured galaxies in our sample. In contrast, far-UV 
attenuation does not appear to depend on star formation activity. These 
results show what kind of studies are possible with {\em CIGALE} for the 
data available for the sample of SINGS galaxies investigated. Since the 
photometric data of the SINGS galaxies is characterised by good coverage of 
the IR and comparatively modest coverage and quality in the UV, we expect 
that for other samples especially at higher redshifts the reliability of the 
different model parameters could be different. For good quality data in the 
rest-frame UV and optical, we can imagine that stellar population properties 
and details of the dust attenuation could be better studied than it has been 
possible for the SINGS sample. On the other hand, the frequently missing 
information in the far-IR for high-redshift galaxies could significantly 
affect the quality of $L_\mathrm{dust}$ and other star-formation-related 
parameters. However, instruments such as {\it Herschel} or ALMA will improve 
the situation in future. Hence, we are convinced that {\em CIGALE} is a 
valuable tool for the characterisation of galaxy populations in the near 
{\em and} distant Universe. In a series of forthcoming papers we will 
demonstrate this by discussing samples of distant galaxies with different 
selection criteria.

\begin{acknowledgements}
SN and DM are funded by the \emph{Agence Nationale de la Recherche} (ANR) of 
France in the framework of the D-SIGALE project. JCMM acknowledges the 
receipt of a Formaci\'on del Profesorado Universitario fellowship from the 
Spanish Ministerio de Educaci\'on y Ciencia, and is also partially financed 
by the Spanish Programa Nacional de Astronom\'{\i}a y Astrof\'{\i}sica under 
grant AYA2006-02358. This publication makes use of data from the {\em Spitzer} 
Infrared Nearby Galaxies Survey (SINGS), the Two Micron All Sky Survey 
(2MASS), the Sloan Digital Sky Survey (SDSS), and the GALEX (Galaxy Evolution 
Explorer) mission. Finally, the authors thank the referee, Adolf Witt, for
his helpful suggestions. 
\end{acknowledgements}

\end{document}